\documentclass[useAMS,usenatbib]{mn2e}
\usepackage{ifthen,graphicx}
\usepackage{epstopdf,aas_macros}
\usepackage{soul,color,amsmath}
\usepackage{rotating,longtable,caption,subcaption}
\usepackage{dblfloatfix}
\def\mathnew{\mathsurround=0pt}
\def\simov#1#2{\lower 2.5pt\vbox{\baselineskip0pt \lineskip-.5pt
\ialign{$\mathnew#1\hfil##\hfil$\crcr#2\crcr\sim\crcr}}}
\def\simless{\mathrel{\mathpalette\simov <}}
\def\simgreat{\mathrel{\mathpalette\simov >}}
\newcommand{\MeV}{Me\kern-0.11em V}
\newcommand{\keV}{ke\kern-0.11em V}

\newcommand{\cha}{{\it Chandra\/}}
\newcommand{\xmm}{{\it XMM-Newton\/}}

\newcommand{\galex}{{\it GALEX \/}}

\newcommand{\ecmss}{erg~cm$^{-2}$ s$^{-1}$}
\newcommand{\es}{erg~s$^{-1}$}
\newcommand{\omegam}{$\Omega_{m,0}$\/}
\newcommand{\omegal}{$\Omega_{\lambda,0}$\/}

\newcommand{\Mbh}{\ensuremath{M_{\bullet}}}

\newcommand{\Rd}{\ensuremath{R_{\rm t}}} 
\newcommand{\Rstar}{\ensuremath{R_{\ast}}} 
 
\newcommand{\Mstar}{\ensuremath{M_{\ast}}} 
\newcommand{\Msun}{\ensuremath{{\rm M}_{\odot}}}

\newcommand{\ciao}{{\it CIAO\/}}
\newcommand{\err}[2]{_{-#2}^{+#1}}
\makeatletter
\newcommand\ion[2]{\mbox{#1$\;${\small\expandafter\@slowromancap\romannumeral #2@\relax}}}
\makeatother
\makeatletter
\newcommand{\raisemath}[1]{\mathpalette{\raisem@th{#1}}}
\newcommand{\raisem@th}[3]{\raisebox{#1}{$#2#3$}}
\makeatother

\title[A Tidal Flare Candidate in Abell 1795]{A Tidal Flare Candidate in Abell 1795\thanks{{\it Herschel} is an ESA space observatory with science instruments provided by European-led Principal Investigator consortia and with important participation from NASA.} \thanks{Based on observations obtained with {\it XMM-Newton}, an ESA science mission 
with instruments and contributions directly funded by 
ESA Member States and NASA}
\thanks{Based on observations made with the NASA Galaxy Evolution Explorer. 
GALEX is operated for NASA by the California Institute of Technology under NASA contract NAS5-98034.}
}
\author[W. P. Maksym, M. P. Ulmer, M. C. Eracleous, L. Guennou, and L. Ho]{W. P. Maksym$^{1,2,3}$\thanks{E-mail:
wpmaksym@bama.ua.edu}, M. P. Ulmer$^{1,2}$, M. C. Eracleous$^{4}$, L. Guennou$^{5}$, \newauthor and L. C. Ho$^{6}$\\
$^{1}$Northwestern University, Department of Physics and Astronomy, Evanston IL 60208, United States of America\\
$^{2}$Northwestern University, CIERA, Evanston IL 60208, United States of America\\
$^{3}$University of Alabama, Department of Physics and Astronomy, Tuscaloosa AL 35487, United States of America\\
$^{4}$The Pennsylvania State University, Department of Astronomy and Astrophysics, State College PA 16802, United States of America\\
$^{5}$Laboratoire d'Astrophysique de Marseille, Marseille 13388, France\\
$^{6}$Observatories of the Carnegie Insititution for Science, Pasadena CA 91101, United States of America}
\begin{document}

\date{Accepted 2013 July 23}

\pagerange{\pageref{firstpage}--\pageref{lastpage}} \pubyear{2012}

\maketitle

\label{firstpage}

\begin{abstract}
As part of our ongoing archival X-ray survey of
galaxy clusters for tidal flares, we present evidence of an X-ray transient source within 1 arcmin of the core of Abell
1795.  The extreme variability (a factor of nearly 50), luminosity ($>2\times10^{42}$ \es), long duration ($>5$ years) and
supersoft X-ray spectrum ($<0.1$ keV) are characteristic signatures of a stellar
tidal disruption event according to theoretical predictions and to
existing X-ray observations, implying a massive $\simgreat 10^5$~\Msun\ black hole at the centre of
that galaxy.  The large number of X-ray source counts ($\sim700$) and long
temporal baseline ($\sim12$~years with \cha\ and \xmm) make this one of the
best-sampled examples of any tidal flare candidate to date.  The transient may be the same EUV source
originally found contaminating the diffuse ICM observations of \cite{BBK99}, which would make it the only
tidal flare candidate with reported EUV observations and implies an early source luminosity 1--2 orders of magnitude greater.  If the host
galaxy is a cluster member then it must be a dwarf galaxy, an order of magnitude
less massive than the quiescent galaxy Henize 2-10 which hosts a massive black
hole that is difficult to reconcile with its low mass.  The unusual faintness of
the host galaxy may be explained by tidal stripping in the cluster core. 
\end{abstract}


\section{Introduction}


If a star passes a massive black hole (MBH) closely enough that its periastron, $R_P$, is less than the tidal radius $R_T\sim\Rstar(\Mbh/\Mstar)^{1/3}$ \citep{Rees88}, the tidal forces may overwhelm the star's self-binding energy and rip it apart in what is commonly known as a tidal disruption event (TDE).  The debris fans out in a long stream, with some fraction of the debris falling back towards the black hole, shocking against the tidal stream, accreting on to the black hole and giving rise to a luminous tidal disruption flare  \citep[TDF,][]{Hills75,YSW77,Young77b,Lacy82,EK89}.  To first approximation the bolometric luminosity should be directly proportional to the mass accretion rate, which is governed by the Keplerian orbits of the debris such that $L=\eta\dot{M}c^2\propto t^{-5/3}$ \citep{Rees88,Phinney89} and emits as a blackbody spectrum that peaks in ultraviolet (UV) or soft X-rays \citep{Ulmer99}.

While TDEs are intrinsically interesting as a specific instance of accretion physics in a relativistic environment, they also have implications for a variety of important astrophysical issues related to the demographics of MBHs and their galactic environments.

An abundance of evidence exists to support the existence of massive black holes (MBHs) at the centres of massive galaxies.  Active galactic nuclei (AGNs) are well-known to provide some of the best evidence, emitting at luminosities that are difficult to explain other than by the sustained accretion of matter on to objects that, due to the Eddington limit, must exceed $10^6$ \Msun.  The determination of the MBH population distribution, particularly at lower ($\la10^6$) masses, is critical to theories of galaxy formation and evolution.  In principle, accretion-based models of black hole evolution \citep{Marconi04} point to an abundance of inactive MBHs in many or most inactive galaxies.  However, aside from our own Galactic centre \citep{Ghez03, Genzel03}, evidence supporting this proposition tends to be indirect and based upon kinematic inferences from spectral modeling of galactic nuclei \citep{Gultekin09b}.

We can infer the low end of the MBH population from the empirical relationship between the mass of a central MBH and the stellar dispersion of its galactic host spheroid \citep[The \Mbh--$\sigma$ relationship, e.g.,][]{Gebhardt00,Gultekin09b}.  Despite extensive work \citep[e.g. using low-mass AGNs,][]{Jiang11,Xiao11}, this mass range remains poorly known, as reliable \Mbh\ estimates for faint, distant dwarf galaxies are difficult to obtain.  AGNs have been identified in low-mass dwarf galaxies such as NGC 4395 \citep{FH03,Peterson05} and POX 52 \citep{Barth04,Thornton08}.  

Improved constraints on the MBH distribution for $\Mbh\la10^6$~\Msun\ in dwarf galaxies would, in particular, help determine the applicability of various scenarios of MBH formation, such as from massive population III star seeds, direct collapse, or runaway stellar mergers in high-redshift clusters \citep[see][for a review]{Volonteri10a}.  Some fraction of dwarf galaxies may harbor only a nuclear star cluster rather than an MBH \citep{FerrareseEtAl06}, while galaxies in groups or clusters may evolve differently due to harassment and more frequent collisions than are typical for field galaxies \citep{Moore96}.  Major mergers could even result in the ejection of the central MBH due to gravitational wave recoil \citep{KM08}.  In addition to constraints on the MBH population, the TDE rate may hold more direct implications for the detection of gravitational waves by any mission similar to the {\it Light Interferometer Space Antenna}\footnote{http://lisa.nasa.gov/} (hereafter
{\it LISA}) which would be sensitive in the mass range of $\Mbh \la 10^{7}\;\Msun$ \citep{Jennrich04,SHMV04,SHMV05,Sigurdsson03,Kobayashi04}.  These implications include the possibility of direct, simultaneous detection of gravitational and electromagnetic signatures from the disruption of a white dwarf \citep{Sesana08}.

Since the first pioneering observational indications of TDEs made by {\it ROSAT} \citep{BKD96,KG99,KB99}, TDEs have been proposed as an explanation for extremely bright extragalactic transients not only in X-rays \citep{Maksym10,Cappelluti09,Esquej08,Esquej07}, but at UV \citep{Renzini95,Cappellari99,Gezari06,Gezari08,Gezari09} and optical \citep{vanVelzen11} wavelengths, as well as in an extragalactic globular cluster \citep{Irwin10}.  Recent developments in tidal flare theory \citep{SQ09,SQ11,LR11} indicate the great potential of optical surveys, such as with Pan-STARRS \citep{PanSTARRS10}, the Catalina Real-time Transient Survey\footnote{http://crts.caltech.edu/}, the Palomar Transient Factory \citep{PTF09} and the Sloan Digital Sky Survey \citep{vanVelzen11} to identify TDEs.  But given many of the difficulties inherent in establishing transient nuclear optical variability \citep{vanVelzen11,Sand08} and disentangling it from that of an ordinary AGN, particularly given the complexity of the early super-Eddington wind emission, X-ray observations remain key to identifying these events and potentially confirming the characteristic $t^{-5/3}$ accretion behaviour.

In our previous paper \citep{Maksym10}, we introduced our ongoing archival X-ray study of galaxy clusters as an attempt to discover new instances of TDFs and to better determine the TDF rate, in particular with respect to the uncertain and possibly dominant dwarf galaxy MBH population \citep{Wang04}.  With their dense galaxy populations, rich clusters have been well-demonstrated to provide an efficient method of locating new transients with a controlled population despite limited fields of view \citep{Maksym10,Cappelluti09,Sand08}.  We also presented an instance of a luminous ($\sim10^{43}$ \es), supersoft ($kT_{BB}\sim0.1$ keV), highly variable (by a factor of over 30) flare best described in terms of a TDF in the galaxy cluster Abell 1689.  Here we describe a second flare with similar X-ray properties but in the direction of Abell 1795 (luminosity distance modulus 37.22, $z=0.062$\footnote{From the NASA/IPAC Extragalactic Database (NED) which is operated by the Jet Propulsion Laboratory, California Institute of Technology, under contract with the National Aeronautics and Space Administration.}) and with a much better-sampled light curve compared to the A1689 flare.  If the galaxy is a cluster member, the observed flare and host galaxy properties imply an exceptional case of an extremely compact ($\sim300$ pc radius) dwarf galaxy flaring from an intermediate-mass ($\ll 10^6$ \Msun) black hole (IMBH) and one of the better examples of X-ray counting statistics reported in a tidal flare to date.  Until we obtain a spectrum to better confirm the likely host galaxy's membership and lack of characteristic AGN emission lines, we must also entertain the alternate explanation of a massive flare from a pre-existing accretion disc.  But in such a case we may have instead identified a similarly exceptional case of a dwarf Seyfert nucleus whose host galaxy is considerably smaller than even such examples as POX 52 and NGC 4395 \citep{Thornton08,Peterson05}.

Throughout this paper, we adopt concordant cosmological parameters of
$H_0=70\ $km$^{-1}$ sec$^{-1}$ Mpc$^{-1}$, \omegam=0.3 and \omegal=0.7,
and calcuate distances using \cite{Wright06}. All coordinates are
J2000.  The galactic column density of neutral hydrogen for Abell 1795
is $1.17\times10^{20}\ $cm$^{-2}$, derived using \cite{Dickey90} values from the {\it
  colden} tool in \ciao\ \citep{ciao06} unless otherwise stated.  All X-ray fluxes and luminosities
used in this paper have been corrected for Galactic absorption using
the assumed column density.


\section{Observations and Data}

\subsection{Overview}

In the course of our galaxy cluster variability survey, we took note of Abell 1795 in particular, due to its excellent temporal coverage by \cha\ and \xmm.  Abell 1795 is moderately rich, with an Abell richness 2 \citep[5 being the maximum,][]{Abell89}.  Since its first \cha\ observations in 1999 \citep{Fabian01}, A1795 has been observed on 17 different epochs by \cha\ using the ACIS camera without gratings.  At $z=0.062$ (1.20 kpc/arcsec), A1795 not sufficiently distant to be completely imaged by ACIS without creation of a mosaic.  Thus, several of the \cha\ observations only partially overlap each other.  The area in which all observations overlap is approximately the size of a single ACIS chip, $\sim8$~arcmin or 0.6 Mpc in diameter, and is centred within $\sim1$~arcmin of the cD galaxy in the region of brightest emission from the intracluster medium (ICM).  A1795 has also been observed once by \xmm\ on 2000 June 26.  These observations are summarized in Table \ref{obs-2}.

\begin{table*}
\centering
\begin{minipage}{140mm}
\caption{X-ray Observations: A1795\label{obs-2}}
\begin{tabular}{lrccccr}
\hline\hline

           &       &             &          & \multicolumn{2}{c}{Aim Point Coordinates} &  \\
\noalign{\vskip -6pt}
           &       & Observation & Duration & \multicolumn{2}{c}{\hrulefill} & Roll \\
Instrument & Obsid & Date (UT)   & (ks)     & $\alpha$(2000) & $\delta$(2000) & Angle \\
\hline

ACIS-S3 & 494 & 1999-12-20 & 19.5 & 13:48:56.382 & +26:36:25.67 & 55.134 \\
ACIS-S3 & 493 & 2000-03-21 & 19.6 & 13:48:49.226 & +26:36:27.27 & 132.79 \\
EPIC & 97820101 & 2000-06-26 & 66.5 & 13:48:53.00 & +26:35:32.00 & 290.15 \\
ACIS-S3 & 3666 & 2002-06-10 & 14.4 & 13:48:48.888 & +26:34:32.27 & 231.72 \\
ACIS-S3 & 5286 & 2004-01-14 & 14.2 & 13:48:55.022 & +26:36:34.94 & 70.118 \\
ACIS-S3 & 5287 & 2004-01-14 & 14.3 & 13:48:55.010 & +26:36:34.97 & 70.236 \\
ACIS-S3 & 5288 & 2004-01-16 & 14.5 & 13:48:54.920 & +26:36:35.36 & 71.182 \\
ACIS-I3 & 5289 & 2004-01-18 & 14.9 & 13:48:55.048 & +26:36:44.95 & 72.664 \\
ACIS-I3 & 5290 & 2004-01-23 & 14.9 & 13:49:00.802 & +26:42:07.52 & 75.924 \\
ACIS-I1 & 6159 & 2005-03-20 & 14.8 & 13:48:32.868 & +26:40:45.42 & 130.98 \\
ACIS-S3 & 6160 & 2005-03-20 & 14.8 & 13:48:49.404 & +26:36:27.11 & 131.08 \\
ACIS-I3 & 6162 & 2005-03-28 & 13.5 & 13:48:47.958 & +26:36:21.74 & 143.86 \\
ACIS-I0 & 6161 & 2005-03-28 & 13.5 & 13:49:19.455 & +26:31:05.40 & 143.27 \\
ACIS-I3 & 6163 & 2005-03-31 & 14.8 & 13:48:47.625 & +26:36:16.42 & 148.56 \\
ACIS-I3 & 10901 & 2009-04-20 & 15.4 & 13:48:48.178 & +26:25:43.11 & 181.54 \\
ACIS-I1 & 10898 & 2009-04-20 & 15.7 & 13:48:46.847 & +26:32:10.44 & 181.88 \\
ACIS-S3 & 10900 & 2009-04-20 & 15.8 & 13:48:47.112 & +26:35:31.44 & 181.23 \\
ACIS-I3 & 10899 & 2009-04-22 & 14.9 & 13:48:21.820 & +26:34:54.59 & 185.55 \\
ACIS-I3 & 12027 & 2010-03-16 & 14.8 & 13:48:35.340 & +26:41:10.87 & 125.17 \\
ACIS-S3 & 12029 & 2010-04-28 & 14.6 & 13:48:47.160 & +26:35:16.87 & 192.29 \\
ACIS-S3 & 12028 & 2010-05-10 & 14.9 & 13:48:47.493 & +26:34:59.68 & 206.00 \\
ACIS-I3 & 12026 & 2010-05-11 & 14.9 & 13:48:46.868 & +26:34:55.14 & 206.51 \\
ACIS-I2 & 13412 & 2011-05-22 & 14.8 & 13:49:18.369 & +26:39:50.08 & 216.55 \\
ACIS-I0 & 13414 & 2011-05-29 & 14.5 & 13:48:59.771 & +26:35:57.80 & 221.89 \\
ACIS-I0 & 13415 & 2011-05-29 & 14.5 & 13:49:18.788 & +26:39:45.59 & 221.93 \\
ACIS-I2 & 13413 & 2011-05-29 & 14.8 & 13:49:16.516 & +26:40:19.97 & 221.67 \\
ACIS-I1 & 13416 & 2011-05-30 & 14.5 & 13:48:41.612 & +26:33:21.55 & 222.22 \\
ACIS-I1 & 13417 & 2011-06-02 & 14.8 & 13:48:32.482 & +26:30:05.47 & 224.60 \\

\hline
\end{tabular}
\end{minipage}
\end{table*}

In the process of applying \cha\ data reduction methods similar to those described in \cite{Maksym10}, including source detection using {\it CIAO wavdetect} and photometry via {\it dmextract}, we manually identified a bright, supersoft temporal drop-out source by examining three band false colour {\it ds9} images of each epoch [bands: S1(0.3--0.9 keV), S2(0.9--2.5 keV), H(2.5-8.0 keV)].  In the earliest epoch, the source was easily visible $\la50$~arcsec northwest of the cluster centre despite having a projected distance of $\sim56$ kpc from the ICM centre.  The source remained visible through at least 2002 June 2010, but was unidentfiable by 2005 March 20 at latest.  These characteristics therefore fit the criteria which we established in \cite{Maksym10} as primary X-ray signatures of tidal disruption flares, and are comparable to those used in previous identifications of TDF candidates \citep{Komossa05,Esquej07}.   By merging the event lists of all available \cha\ epochs with {\it merge\_all} we obtained the net (i.e. background-subtracted) number of source counts for all observations using {\it wavdetect} within the central 5~arcmin on the S1, S2 and H bands separately.  Despite only being bright in the earliest observations, the source had 1128.8 net S1 counts over all epochs.  The next brightest S1 objects were two peaks in ICM at $\sim3$~arcsec to the northwest of the cD galaxy, with net S1 counts of 268.3 and 295.1 for radii of $\sim2.1$~arcsec and $\sim1.2$~arcsec respectively.  No other source in the field had greater than 180 net S1 counts.  Despite the source's brightness and proximity to the ICM centre, it receives no attention in \citep{Fabian01}, nor (to the best of our knowledge) any specific mention in later X-ray studies of the cluster.

These characteristics immediately identify the source as an X-ray transient, even in the absence of the detailed population variability analysis which we applied in our previous examination of Abell 1689 \citep{Maksym10} and will apply to Abell 1795 in a subsequent paper.  

In order to determine the suitability of this A1795 flare as a TDF candidate and compare against other variable X-ray sources such as AGNs, supernovae (SNe), Gamma Ray Bursts (GRBs) and flaring M-dwarf stars, we have undertaken X-ray variability and spectral analysis of A1795 using the available \xmm\ and \cha\ data.  In addition, we have examined archival optical, ultraviolet and infrared data from {\it HST WFPC2}, SDSS \citep{SDSSdr7} and the WINGS survey \citep{Varela09}, {\it GALEX}, and {\it Spitzer}.  We will show that correlation of this supporting multi-wavelength data with the X-ray point source position demonstrates that it is associated with a faint ($V\sim22.5$) galaxy.  In addition, given the flare was at its brightest in the earliest stages of the \cha\ mission (1999 December 20), we examined seven {\it EUVE} observations obtained between 1997 February 3 and 1999 May 31 to determine if earlier emission could be detected in the anticipated maximal range of blackbody emission for a tidal flare.  We find a possible association with a luminous flare first reported by \citep{BBK99}.

\subsection{X-ray Observations}  

\subsubsection{Source Position}

To determine the source position, we used the {\it wavdetect} tool from {\it CIAO} 4.0.2 \citep{ciao06} on an 8.5-arcmin~$\times$~8.5-arcmin image from the 1999 December 20 archival \cha\ L2 event file.  We centred the image at $(\alpha,\delta)=(13^{h}48^{m}51^{s}.1$, $+26\degr35\arcmin05\arcsec.7 )$ (J2000) to include other obvious bright sources with one spatial bin per pixel.  The $wavdetect$ position for this source was $(\alpha,\delta)=(13^{h}48^{m}49^{s}.86$, $+26\degr35\arcmin57\arcsec.49 )$ (J2000).  Multiple detected X-ray sources corresponded to SDSS galaxy positions within one ACIS pixel ($\sim0.5$~arcsec), indicating comparable absolute astrometric accuracy.  The point spread function (PSF) radius for $39.3$ per cent encircled energy is 0.52~arcsec at the source's off-axis location, as determined by {\it wavdetect}.

\subsubsection{XMM detection}

The \xmm\ data pipeline failed to detect a point source within the \xmm\ PSF ($\sim20$-arcsec).  However, we examined a three colour false image from the pipeline products (0.2-0.5 keV, 0.5-1.0 keV, 1.0-2.0 keV) and found a possible extremely soft source cospatial with the flare identified via \cha\ data, embedded in the diffuse ICM emission.  To confirm the source, we used {\it wavdetect} on the archival multi-band images but failed to detect a source at the flare position.  Using {\it evselect} from {\it XMM-SAS}\footnote{http://xmm.esac.esa.int/sas/}, we produced another image filtered between 0.2 and 0.5 keV, as per the pipeline products, but with 40 pixels per spatial bin rather than the default value of 80 used by \xmm\ pipeline processing.  This method detected a source significant to 18.5$\sigma$ and within 1.0~arcsec of the \cha\ source (compared to $\sim1.5$-arcsec pointing accuracy and $\sim20$-arcsec PSF).  There is no obvious ($\simgreat 3\sigma$) detection with the UVOT in the UVW1, UVW2, UVM1, UVM2 or U bands, however the field suffers from strong ghost images due to the reflection of off-axis bright sources at the cluster centre in the area of the X-ray source. 

\subsubsection{Flare Photometry}

\begin{figure}
\includegraphics[width=3.15in]{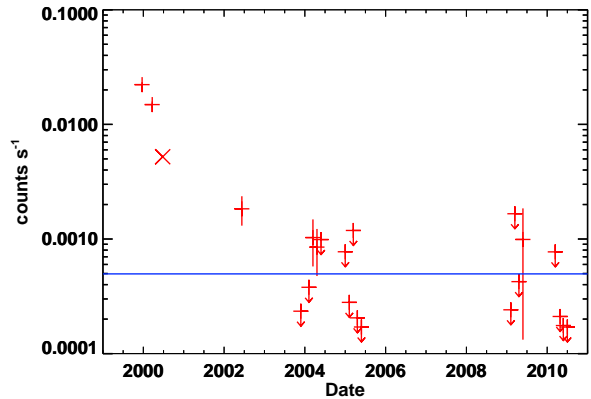}
\caption[\cha\ Count Rate Evolution for WINGS~J1348]{\cha\ count rate evolution for WINGS~J1348.   Red crosses indicate soft (0.3--2 keV) X-rays.  Count rate uncertainty is indicated by cross extent.  Arrows indicate $2\sigma$ upper limits.  The $\times$ is a \cha\ ACIS-S equivalent count rate derived from the \xmm\ epoch via its best-fitting power law model.  After 2004, individual observations within a given year are offset along the X-axis for clarity, and would otherwise overlap.  The blue horizontal line is the median $2\sigma$ upper limit in the hard (2--8 keV) band.}
\label{counts_lc}
\end{figure}

Inspection of \cha\ images revealed that the source remains bright through 2002 June 10 but afterwards becomes difficult  in individual observations to distinguish from statistical fluctuations in the local ICM.  Using a 95 per cent encircled energy extraction radius for the \cha\ PSF at 0.5~keV and background annulus covering 4 times the extraction region, we use {\it dmextract} to derive count rates for S1, S2, S, H and B(0.3--8.0 keV) bands.  The time evolution of the \cha\ count rate is indicated in Figure \ref{counts_lc}, with upper limits for non-detections.  Note that because we have assumed a 95 per cent encircled energy extraction radius, actual detections are possible for some later (2004--2005) epochs with which have only $2\sigma$ upper limits indicated.  With large off-axis angles, as a 95 per cent encircled energy extraction radius includes more of the bright central diffuse emission for a source close to the cluster core and hence infers a larger background rate.

\subsubsection{Spectral Fitting and Evolution}\label{specfitsec}

We fitted various spectral models for separate epochs of the flare using {\it XSPEC} v12.7.1 \citep{Arnaud96,Dorman01}.  As in \cite{Maksym10}, we used \cite{chur96} weighting and included {\it colden} neutral hydrogen ($N_H$) extrapolations from \cite{Dickey90}, assuming negligible redshift effects.  To reduce the contribution of the strong diffuse intracluster emission, we use 90 per cent encircled energy extraction radii for spectral fitting, calculating \cha\ spectra and position-dependent response matrices using {\it specextract}.  

For \cha\ spectra, the background is determined from a concentric circular annulus between the source extraction radius and 2.5 times that radius.  These annuli are typically small ($\sim3$-arcsec radius), so background varies modestly across the extraction region ($\sim25\%$ variation from the mean value, consistent with Poisson statistics) and is likely representative of the source region.  According to \cite{Gu2012}, the ICM in this region is stable to $\sim0.5\;$keV in projected temperature, and to $\sim0.1\;$solar in abundance.  The \cha\ background flux prior to Epoch 5 (Jan. 2004) is $<4\%$ of the source flux in all cases, and therefore within the uncertainties in the fit parameters.  Due to the low source flux, epoch 5 is therefore the only \cha\ epoch whose fits may be significantly affected by improper background subtraction.  The \xmm\ spectrum requires more careful treatment due to the larger PSF, and is addressed later in this section.

Where fitting would bring $N_H$ unphysically low, we froze it at the {\it colden} value.  Prior to 2004, we treat each observation individually as an epoch.  After 2004, however, the source count significance of any individual observation is so low that a meaningful fit is impossible.  For these later observations, we merge spectra from multiple observations over a given annual observing cycle into a single `epoch' in order to fit an average spectral model over typical timespans of weeks.  

The results of these fits are presented in Table \ref{XrayFits}, which covers the soft (0.2--2.0 keV) band that contains the vast majority of photons detected over the ICM (as illustrated via the hard and soft count rates in \ref{counts_lc}), and in Table \ref{XrayHard}, which covers the hard (2.0--8.0 keV) band which has negligible source counts.  

\begin{table*}
\centering
\begin{minipage}{100mm}
\caption{WINGS~J1348 X-ray Spectral fits: 0.2--2.0 keV\label{XrayFits}, Epochs 1-9}
\begin{tabular}{@{ }lrccccr@{}}

\multicolumn{2}{c}{Emission Model} & 					&			&	\\
 \noalign{\vskip -6pt}
 \multicolumn{2}{c}{\hrulefill} 		& 					&			&Model Flux	\\
Power Law	&	Blackbody	&	object $N_H$		& $\chi^2$/dof	& (68\% conf. error) \\
$\Gamma$	&	kT (keV)		&	$10^{20}$ cm$^{-2}$	&			& $10^{-14}$ \ecmss	\\
\hline 

\hline 
\multicolumn{5}{c}{Epoch 1: \cha, 1999-12-20, obsid 494}\\
\hline 
$4.21\pm 0.11$& -				& $<0.08	$		&	124.77/119	& $20.6\err{2.3}{2.2}$\\ 
4.21*		& -				& 0.00*			&	124.77/121	& $^\dagger20.8\err{1.0}{0.9}$\\
-			& $0.084\pm0.003$	& $<0.02	$		&	174.73/119	& $13.3\err{2.3}{2.9}$\\ 
-			& 0.08*			& 0.00*			&	175.57/121	& $^\dagger13.8\err{0.6}{0.6}$\\
-			& $^{bb}0.025^*,0.106$		& $<0.165$		&	120.88/118	& $^\dagger29.5\err{2.1}{2.0}$\\
-			& $^{d1}0.103\pm0.007$ & $<2.07$			&	165.89/121&$14.2\err{1.7}{4.5}$ \\
-			& $^{d2}0.099\pm0.004$ & $0.00$			&	156.70/119	&$14.3\err{0.7}{0.9}$ \\
-			& $^{c1}0.064\pm0.003$ & $0.00$			&	146.12/118	& $ 14.8\err{2.7}{9.7}$ \\
-			& $^{c2} 0.011\pm0.003$ 	& $16.1\pm9.5$	&	{\bf 117.02/117}& $31.4\err{0.1}{0.1}$ \\
\hline 
\multicolumn{5}{c}{Epoch 2: \cha, 2000-03-21, obsid 493}\\
\hline 
$4.85\pm0.33$	& -				& $3.04\pm1.31$	&	133.62/119	& $29.4\err{10.6}{8.4}$\\ 
4.21*		& -				& $0.74\pm0.53$	&	{\bf 131.15/120}	& $17.1\err{1.7}{1.8}$\\ 
4.21*		& -				& 0.00*			&	139.88/121	& $^\dagger14.4\err{0.8}{0.8}$\\
-			& $0.090\pm0.006$	& $<2.82$			&	138.41/119	& $9.6\err{1.9}{2.2}$\\ 
-			& 0.08*			& 0.00*			&	143.74/121	& $^\dagger10.4\err{0.6}{0.5}$\\
\hline 
\multicolumn{5}{c}{Epoch 3: \xmm, 2000-06-26, obsid 0097820101}\\
\hline 
$5.55\pm1.33$ & -				& 0.00*			& 	168.43/134	& $<5.17$\\
$5.95\pm1.28$ & -				& 1.00*			&	167.84/134	& $<7.43$\\
$7.82\pm1.59$ & -				& 5.00*			&	166.05/134	& $<32.6$\\
$9.50\pm5.08$ & - 				& 10.00*			&	165.51/134	& $<143$\\
4.21*		& -				& $<2.28$			&	171.58/135	& $4.4\err{2.6}{2.4}$\\
4.21*		& -				& 0.00*			&	171.58/136	& $^\dagger4.4\err{0.9}{1.0}$			\\
- 			& $<0.43$			& $<0.48$			&	{\bf 163.69/134}	& $<15.6$		\\
-			& 0.024*			& 0.00*			&	175.22/136	& $^\dagger6.0\err{1.5}{1.5}$\\
-			& 0.08*			& 0.00*			&	171.61/136	& $^\dagger3.26\err{0.7}{0.7}$\\
$3.05\pm0.24$	& -				& $<1.08$			&	228.48/181	& $^\ddagger13.5\err{2.5}{2.5}$\\
-			& $0.105\pm0.006$	& $<0.54$			&	262.20/181	& $^\ddagger10.1\err{1.7}{1.7}$\\
\hline 
\multicolumn{5}{c}{Epoch 4: \cha, 2002-06-10, obsid 3666}\\
\hline 
$5.03\pm1.90$	& -				& $<5.49$			&	171.47/119	& $3.9\err{5.1}{3.4}$\\ 
$5.15\pm0.61$	& -				& 0.00*			&	171.44/119	& $4.0\err{3.6}{1.8}$\\ 
4.21*		& - 				& $<2.97$			&	{\bf 169.30/120}	& $3.4\err{1.1}{1.1}$\\ 
4.21*		& -				& 0.00*			&	174.31/121	& $^\dagger3.0\err{0.6}{0.6}$\\
-			& $0.057\pm0.022$	& $<12.8$			&	172.61/119	& $<4.2$\\ 
-			& $0.057\pm0.007$  & 0.00*			&	172.61/119	& $^\dagger3.1\err{0.4}{0.9}$\\
-			& 0.08*			& 0.00*			&	179.54/121	& $^\dagger1.9\err{0.4}{0.4}$		\\
\hline
\multicolumn{5}{c}{Epochs 5-9: {\it continued on next page}}\\
\hline
\end{tabular}

* frozen\\
$\dagger$ flux derived when all parameters except normalization are frozen and then refit.  All models are redshifted to 0.062 and assume galactic absorption of
$N_H=1.17\times10^{20}$ cm$^{-2}$.  $\chi^2$/dof for best fits is {\bf bold} prior to 2005. \\
$\ddagger$ background is modeled according to a two-component APEC model, as described in the text.\\
$bb$ This model is the sum of two independent blackbody models, as in \S\ref{specfitsec}.
$d1$ {\tt diskbb}, with $kT$ corresponding to the temperature of the inner disc.  Normalization = $130.8\pm66.1$.\\
$d2$ {\tt ezdiskbb}, with $kT$ corresponding to the temperature of the inner disc.  Normalization = $26.0\pm6.0$.\\
$c1$ {\tt compbb}, with electron temperature frozen at 50 keV and plasma optical depth $\tau=0.188\pm0.038$, normalization $= 1563\pm425$.\\
$c2$ {\tt compbb}, with electron temperature $kT_e=15.8\pm2.4$, plasma optical depth $\tau<0.002$, normalization $< 3.1\times10^{15}$.
\end{minipage}
\end{table*}

\begin{table*}
\centering
\begin{minipage}{100mm}
\contcaption{\\ WINGS~J1348 X-ray Spectral fits: 0.2--2.0 keV, Epochs 1-9}
\begin{tabular}{@{ }lrccccr@{}}

\multicolumn{2}{c}{Emission Model} & 					&			&	\\
 \noalign{\vskip -6pt}
 \multicolumn{2}{c}{\hrulefill} 		& 					&			&Model Flux	\\
Power Law	&	Blackbody	&	object $N_H$		& $\chi^2$/dof	& (68\% conf. error) \\
$\Gamma$	&	kT (keV)		&	$10^{20}$ cm$^{-2}$	&			& $10^{-14}$ \ecmss	\\
\hline 
\multicolumn{5}{c}{Epoch 5: \cha, 2004-01-14 to 2004-01-23, obsids 5286-5290}\\
\hline 
$6.29\pm3.25$	& -				& $<15.9$			&	{\bf 95.13/120}		& $<0.70$		\\
$6.33\pm2.74$	& -				& 0.00*			&	{\bf 95.13/120}		& $<0.73$		\\
4.21*		& -				& 0.00*			&	96.76/121		& $^\dagger0.89\err{0.25}{0.25}$\\
-			& $0.035\pm0.109$	& $0.2\pm54.9$	&	96.32/120		& $^\dagger1.73\err{0.42}{0.46}$		\\
-			& 0.08*			& 0.00*			&	101.68/121	& $^\dagger0.61\err{1.9}{2.27}$\\
\hline 
\multicolumn{5}{c}{Epoch 6: \cha, 2005-03-20 to 2005-03-31, obsids 6159-6163}\\
\hline 
4.21*		& -				& 0.00*			&	113.82/121	& $^\dagger<0.924$			\\
-			& 0.08*			& 0.00*			&	113.78/121	& $^\dagger<0.489$			\\
\hline 
\multicolumn{5}{c}{Epoch 7: \cha, 2009-04-20 to 2009-04-22, obsids 10898-10901}\\
\hline 
4.21*		& -				& 0.00*			&	120.52/121	& $^\dagger<0.603$			\\
-			& 0.08*			& 0.00*			&	120.52/121	& $^\dagger<1.044$			\\
\hline 
\multicolumn{5}{c}{Epoch 8: \cha, 2010-03-16 to 2010-05-11, obsids 12026-12029}\\
\hline 
4.21*		& -				& 0.00*			&	110.90/121	& $^\dagger<0.441$			\\
-			& 0.08*			& 0.00*			&	111.06/121	& $^\dagger<0.354$			\\
\hline 
\multicolumn{5}{c}{Epoch 9: \cha, 2011-05-22 to 2011-06-02, obsids 13412-13417}\\
\hline 
4.21*		& -				&0.00*			&	109.65/121	& $^\dagger<7.752$			\\
-			& 0.08*			&0.00*			&	109.65/121	& $^\dagger<5.551$			\\
\hline 


\end{tabular}

* frozen\\
$\dagger$ flux derived when all parameters except normalization are frozen and then refit.  All models are redshifted to 0.062 and assume galactic absorption of
$N_H=1.17\times10^{20}$ cm$^{-2}$.  $\chi^2$/dof for best fits is {\bf bold} prior to 2005.
\end{minipage}
\end{table*}

\begin{figure}
\includegraphics[width=3.15in]{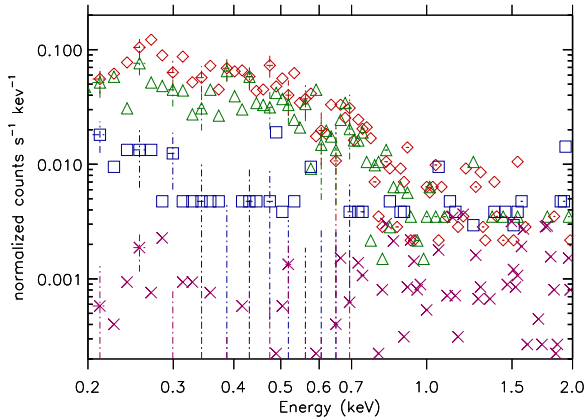}
\caption[Overlaid \cha\ background-subtracted spectra (0.2--2.0 keV) of WINGS J1348 at early epochs]{Overlaid \cha\ background-subtracted spectra (0.2--2.0 keV) of the WINGS J1348 TDFC at early epochs, where photon counts are sufficient to fit free parameters beyond the normalization.  Top (red diamonds) is 1999 December 20.  Second (green triangles) is 2000 March 21.  Third (blue squares) is 2002 June 10.  Bottom (magenta $\times$) is 2004 January 14.  Epoch data are connected by dashed lines.  To prevent crowding, only every third error bar (dot-dash lines) is indicated, and all error bars above 0.7 keV (which are comparable to those at 0.7 keV) are omitted.  The decay in flux is almost monotonic at energies below $\sim0.8$ keV, while at higher energies all epochs are almost indistinguishable from the diffuse ICM background.}
\label{specov}
\end{figure}

The peak epoch, CXO1, is a moderately good fit to both a steep ($\Gamma\sim4.21$) power law and an extremely soft ($kT=0.084$ keV) blackbody fits, although the power law fit is significantly better in CXO1.  All epochs similarly well-fit best steep ($\Gamma>4$) power laws or soft ($kT<0.1$ keV) blackbodies.  For later epochs where detections are marginal or nonexistent, we fit determine the soft X-ray flux $F_X(0.2-2.0~\rm{keV})$ with $\Gamma$ or $kT$ comparable to the best-fitting values of the early epochs.  The X-ray spectrum declines monotonically at all energies, as can be seen in Fig. \ref{specov}.  

The fits of these simple models are complicated by the extreme softness of the source, such that the sharp quantum efficiency cutoff of ACIS-S becomes a significant issue, as well as photon energy resolution near the C-K edge at 0.285 keV and the time evolution of contaminant buildup on ACIS that blocks a significant fraction of photons incident on the detector \citep{ACIScontam}.  As the instrument is not well-calibrated below 0.25 keV, for \cha\ fits we only consider power law data fit between 0.25 keV and 2 keV, above which the background strongly dominates the source.  Soft band fluxes, $F_X$(0.2--2.0 keV), are calculated by extrapolating of the model beyond the \cha\ lower bound, and are corrected for galactic absorption.

The difficulty of extrapolating the source flux to low energies is compounded by excess photons above both power law and blackbody fits at energies approaching 0.2 keV.  This observed excess is difficult to explain solely by evolution of the ACIS contaminant, as the diffuse ICM in epoch 2 has only $\sim3\%$ fewer photons at 0.2--0.3 keV within $\sim0.5\arcmin$ of the cluster core compared to 0.5--0.7 keV.  WINGS J1348, on the other hand, has $\sim21\%$ fewer photons in 0.2--0.3 keV in epoch 2 compared to monotonic decline, or $\sim2\sigma$ below the expected value.

The addition of any component to account for this apparent excess may result in large variations in estimates of the bolometric luminosity $L_{bol}$.  As we are considering a tidal disruption event as an explanation for this flare, we also consider a two-blackbody fit for energies down to 0.20 keV during CXO1.  This model is a rough approximate of two physically plausible characteristic radii for tidal flare emission, namely the shocked material at the tidal disruption radius \Rd\ and the innermost edge of the accretion disc near the Schwarzschild radius $R_S$ or the innermost stable circular orbit $R_{ISCO}$ \citep{Ulmer99}.  We observe that the source is well-fit ($\chi/\nu$=120.88/118) where ($kT_1=0.025$ keV, frozen) and ($kT_2=0.105\pm0.023$ keV), significantly better than any of the blackbody or power law fits.

In order to consider other physically motivated scenarios, we have also fit CXO1 to the {\tt diskbb} \citep{diskbb84,diskbb86} and {\tt ezdiskbb} \citep{ezdiskbb05} multi-colour blackbody disc models, as well as the {\tt compbb} Comptonised blackbody model \citep{compbb86}.  At $z=0.062$, the normalization of {\tt diskbb} implies $\Mbh=1.06\err{0.24}{0.31}\times10^5\;(R_{ISCO}/R_{S})^{-1}{(cos\;i)}^{-1/2}\;$\Msun, where $i$ is inclination;  {\tt ezdiskbb}, which imposes a zero-torque condition and is suited to thin discs where the radiation is emitted at the ISCO, implies $\Mbh=1.41\err{0.13}{0.21}\times10^5\;f^{2}{(cos\;i)}^{-1/2}\;$\Msun, where $f$ is the ratio between colour temperature and effective temperature in the disc.  As with blackbody fits, the disc models fit to CXO1 leave an excess of $\sim0.04$~counts~s$^{-1}$~cm$^{-2}$ at 0.2--0.3 keV.  The bolometric corrections for these models are $\sim2$.

The normalization of {\tt compbb} requires a relatively large bolometric luminosity, $L_{bol}=1.10\err{0.14}{0.16}\times10^{45}\;$\es\ at the default electron temperature $kT_e=50\;$keV, or an essentially unconstrained $L_{bol}<3.8\times10^{48}\;$\es\ for the best-fit $kT_e=13.7\pm5.9\;$keV, $\tau<1.18$.   Furthermore, because the bulk of the blackbody energy in the {\tt compbb} model can be produced below the energy range at which \cha\ detects photons, such that the data primarily sample the Comptonised tail.  Thus, a wide range of statistically significant (null hypothesis $p>0.05$) minima exist for {\tt compbb} in addition to the best-fit parameters in Table \ref{XrayFits}, which generally require $L_{bol}\simgreat10^{45}\;$\es, $N_H\simless10^{21}\;\rm{cm}^{-2}$, $kT\simless0.06$, $kT_e\simless50\;$keV, and $\tau\simless1$.

For all blackbody and power law models, there are variations in the data from the best-fit continuum on small ($\sim0.1$ keV width) scales, particularly between 0.6 and 0.9 keV in the observer frame. 
  This may be interpreted as line or edge absorption at either end of this range, or as an emission line near 0.68 keV.  Further detail, however, is beyond the scope of this analysis.

We also fit a power law ($\Gamma=2$) to place an upper limit on $F_X$(2.0--8.0 keV) for all \cha\ and \xmm\ epochs, assuming a spectrum comparable to a typical AGN (Table \ref{XrayHard}).  Most limits set by \cha\, with its small PSF relative to \xmm, are below $8\times10^{-15}$ \ecmss, and reach $5.1\times10^{-15}$ \ecmss\ in the earliest epoch.\\

\noindent{\bf XMM Spectrum:} For EPIC PN data from the \xmm\ epoch, we used the standard {\it SAS}\footnote{http://xmm.esac.esa.int/sas/} v7.1.0 spectral extraction tools {\it evselect}, {\it arfgen} and {\it rmfgen}.  The \xmm\ spectrum is extracted as a point source from a 15-arcsec region.  The size of the \xmm\ PSF is significant relative to the ICM core (FWHM$\sim27$~arcsec) and the source separation from the ICM peak ($\sim50$~arcsec), incorporating $\simgreat2\times10^4$ background counts.  For fits which follow the same methodology as \cha\ data, we therefore choose an adjacent background extraction circle of similar size and separation from the cluster core, with nearly identical position with respect to the $0.1-12\;$keV isophotal contours of the ICM,  $(\alpha,\delta,\rho)=(13^{h}48^{m}49^{s}.3$, $+26\degr35\arcmin26\arcsec.1,15\arcsec)$.  This choice of background region minimizes the effect of the ICM on the total number of background counts, as well as $F_X$(0.2--2 keV) for data which are fit only below 2 keV.

The spectrum of the background extraction region may vary spatially in ways which significantly affect our results, however.  The deprojected ICM of A1795 in this region may be characterized as a two-component ($\sim2\;$keV and $\sim6\;$keV) APEC model, and the background extraction region may have modestly cooler temperature than the extraction region, as much as $\sim1\;$keV according to a single-temperature ICM model \citep{Gu2012}.  We therefore test the validity of our background-subtracted fit by independently modeling the ICM in the source extraction region.  We assume that emission above 1 keV is strongly dominated by the ICM, and fit it to an absorbed two-component APEC model \citep[galactic $N_H$, $kT_1=2.51\;$keV,  $kT_2=5.88\;$keV, abundance $A=0.74$, as per][for \cha\ at 30--51 $h^{-1}_{71}$ kpc, with all parameters frozen except normalization]{Gu2012}.  We then add an absorbed blackbody or power law (representing the point source component) to fit the 0.1--9 keV spectrum.  The results of these fits are included in Table \ref{XrayFits}.  Note that this method results in a significantly harder source spectrum, and tends to minimize the intrinsic column density.  This method also produces the blackbody temperature most consistent with those derived from \cha\ epochs.

We do not subtract a {\it Suzaku}-detected $\sim0.8\;$keV component to the ICM  found by \cite{Gu2012} within $\sim144\;h^{-1}_{71}$ kpc of the cluster core (approximately the limit of {\it Suzaku}'s angular resolution).  If the component were uniformly distributed over the region, we would expect a contribution of $\sim6.0\times10^{-15}\;$\ecmss\ within the \xmm\ source region.  

If, instead, the soft ICM component were associated with the central cooling filament \citep[$\sim0.7\%$ of the total core luminosity, as per][]{Gu2012}, then its flux contribution to the \xmm\ extraction region would still be insignificant given the $\sim41$--arcsec separation between the filament and the source.

\begin{table*}
\centering
\begin{minipage}{150mm}
\begin{center}
\caption{X-ray Spectral Fits: 2.0--8.0 keV Upper Limits\label{XrayHard}}
\begin{tabular}{llcc}
\hline
		&				&				&	$F_X$(2.0--8.0 keV) \\
		& Epoch			&	$\chi^2$/dof	&	(68\% conf. limit) \\
		& 				&				&$10^{-14}$ \ecmss \\
\hline
1: \cha, & 1999-12-20, obsid 494						& 175.74/409	&	$<0.51$ \\
2: \cha, & 2000-03-21, obsid 493						& 79.5/409	&	$<0.78 $\\
3: \xmm, & 2000-06-26, obsid 0097820101				& 66.73/409	&	$<3.99 $\\
4: \cha, & 2002-06-10, obsid 3666					& 50.79/39		&	$<0.69$ \\
5: \cha, & 2004-01-14 to 2004-01-23, obsids 5286-5290				& 314.21/409		&	$<0.67$	\\
6: \cha, & 2005-03-20 to 2005-03-31, obsids 6159-6163					& 297.26/409		&	$<1.11 $\\
7: \cha, & 2009-04-20 to 2009-04-22, obsids 10898-10901				& 337.33/409	&	$<1.78 $\\
8: \cha, & 2010-03-16 to 2010-05-11, obsids 12026-12029				& 300.80/409		&	$<0.75 $\\
9: \cha, & 2011-05-22 to 2011-06-02, obsids 13412-13417				& 353.17/409	&	$<1.80 $\\

\hline
\end{tabular}
\end{center}
All fits in this table assume galactic absorption $N_H=1.17\times10^{20}$ cm$^{-2}$, negligible intrinsic absorption, and a power law $\Gamma=2.0$ redshifted to $z=0.062$ fit between 2.0 and 8.0 keV.  The only free parameter is normalization.
\end{minipage}
\end{table*}

\noindent{\bf Hardness Evolution:}  Although the spectra are reasonably well-fit by various models, they do not appear to be uniquely constrained by any given choice of model.  We therefore examine the evolution of spectral hardness in a more model-independent way.  The \cha\ and \xmm\ effective areas are strongly energy-dependent below 2 keV, which is also the regime where almost all photons are detected.  In addition, the \cha\ instrumental response varies strongly as a function of time, due to the aforementioned contaminant build-up.   Hardness ratios must therefore be corrected for instrumental effective area in order to be useful.  We use a {\it Bayesian Estimation of Hardness Ratios} tool \citep[BEHR;][]{BEHR06} with default priors for bands defined at 0.2--0.5~keV for the soft input, and 0.5--1.0~keV for the hard input.  Given the low flux at higher energies (see, e.g., Table \ref{XrayHard}), harder bands do not produce useful information and are excluded.

\begin{figure}
\includegraphics[width=3.15in]{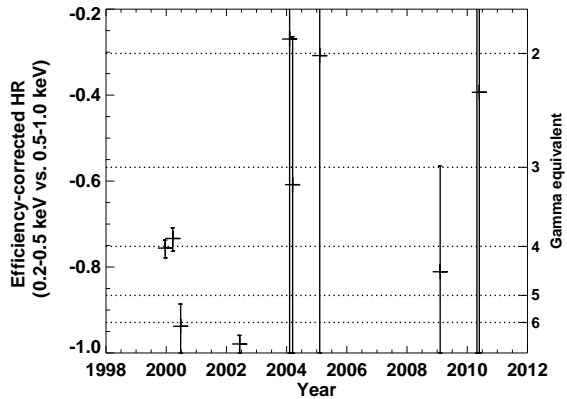}
\caption[Efficiency-corrected HR Evolution]{Evolution of the hardness ratio \citep[HR, as defined in][]{BEHR06} between the 0.2--0.5~keV and 0.5--1.0~keV bands as a function of time, as determined using BEHR, with epoch-dependent effective area calculated as for a power law with $\Gamma=4.21$ and galactic absorption.  Horizontal dotted lines indicate simulated hardness ratios for power laws with $\Gamma=(2,3,4,5,6)$, galactic absorption, and total number of counts as per CXO1.  Only pointings with effective areas greater than 100 cm$^2$ at 0.5--1.0~keV are indicated. HR=0 when the ratio between net count rates is equal to the ratio between effective areas in the respective bands.}
\label{hardness}
\end{figure}

Hardness evolution as a function of time is plotted in Fig. \ref{hardness}.  Even within these relatively narrow supersoft bands, the source is quite soft for all epochs with meaningful constraints, and it appears to soften sharply by the 2000 \xmm\ epoch, and remain soft at least through 2002, as $F_X(0.2-2.0)$ decreases.

\subsection{Supplementary Observations}

\subsubsection{HST/WFPC2 Observations}

\begin{figure*}
\centering
\begin{subfigure}{3.in}
\includegraphics[width=2.7in,trim=2in 0in 1.5in 0in,clip]{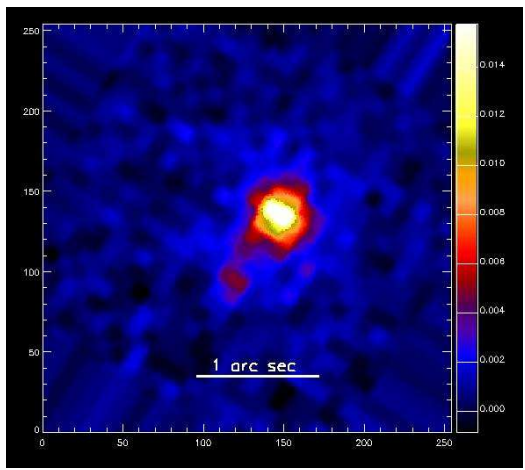}
\end{subfigure}
\qquad
\begin{subfigure}{3.in}
\includegraphics[width=2.7in,angle=90]{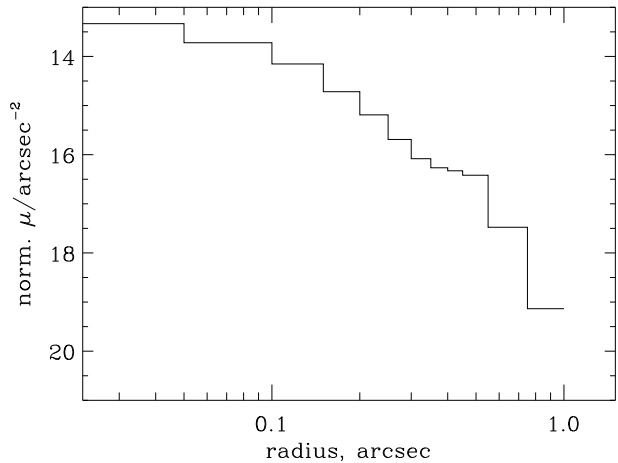}
\end{subfigure}
\caption[HST Image and Radial Profile of WINGS~J1348]{Left: WFPC2 image of WINGS~J1348, from F555W and F814W images combined to remove cosmic rays.  Right: Solid line, radial Profile of WINGS~J1348.  The y-axis is in units of magnitudes per square arcsecond for an annulus of mean radius corresponding to the x-axis.}
\label{hstdata}
\end{figure*}

The \cha\ source had been observed by the {\it Hubble Space Telescope} ({\it HST}) on 1999 April 22 using WFPC2 for 300 s each in the F555W and F814W filters.  By matching with known Sloan Digital Sky Survey (SDSS) objects within the WFPC2 field of view, we corrected an absolute astrometry offset and found a small ($\sim 0.3$-arcsec) extended object, larger than the {\it HST} PSF but within the \cha\ $\sim0.52$-arcsec PSF.  Images were processed using the {\it MultiDrizzle} tool, and cosmic rays were removed by assuming similar photometric profiles for both F555W and F814W bands, then interpolating the contaminated regions by rescaling pixel values from the locally uncontaminated {\it HST} band.  The radial photometric profile from the averaged band images, as in Fig. \ref{hstdata}, is broader than the 0.1-arcsec resolution of WFPC2.  The profile width confirms the object is extended, and at $>75$\degr\ galactic latitude likely a galaxy.  The F814W magnitude must be treated with caution, as a cosmic ray is within $\sim0.5\;$arcsec of the object center, but upon removing the cosmic ray we infer $F814W=21.5\pm0.4$ using {\it ATV} aperture photometry \citep{ATV}.

\subsubsection{WINGS and SDSS photometry}\label{WSphot}

SDSS is sensitive to $u\sim22.0$, $g\sim22.2$, $r\sim22.2$, $i\sim21.3$, $z\sim20.5$, and the source position was observed on 2004 June 13.  There is, however, no object within 3~arcsec of the source position in the SDSS DR 7 catalog.  A faint object is however visible in an SDSS finding chart centred on that position, falling presumably just below the detection limit of the survey.  The SDSS observations are consistent with {\it HST} results showing $F814W\sim21.5$, just below the detection limits of $i,z$.  An object does appear in the WIde-field Nearby Galaxy Cluster Survey (WINGS) \citep{Varela09}.  WINGS J134849.88+263557.5 (hereafter WINGS J1348) is classified by that survey as a galaxy, and is detected at $V=22.46$ and $B=23.28$ isophotal magnitudes via SExtractor \citep{SExtractor96}.  WINGS fails to detect an object in the $J,K$ and is 90 per cent complete to $J\sim20.5$, $K\sim19.4$ \citep{Valentinuzzi09}.  Due to photometric uncertainty introduced in {\it HST} data in our method of cosmic ray removal, we prefer the deep WINGS values in subsequent analysis, but note that the magnitude inferred from F555W is $\simless 10\%$ different from the WINGS $V$ value.

\subsubsection{GALEX, Spitzer and Herschel Non-detections}

The location of WINGS J1348 has been observed twice during the All-Sky Survey of the {\it Galaxy Evolution Explorer}\footnote{http://www.galex.caltech.edu}.  On 2005 April 18 at 23~arcsec from the aimpoint, the object was undetected to $NUV\sim23$ ($\sim1500$\AA\ ) during 300~s, and in $FUV\sim22$ ($\sim2300$\AA) during 100~s.  WINGS J1348 was also undetected on 2007 May 19 at 34~arcsec from the aimpoint over 100 s in both $NUV$ and $FUV$.

WINGS J1348 was also observed by the {\it Spitzer}\footnote{http://www.spitzer.caltech.edu} MIPS at 24$\mu$m, 70$\mu$m, and 160$\mu$m wavelengths on 2004 July 11, and again with {\it Spitzer} IRAC at 4.5$\mu$m, and 8.0$\mu$m on 2010 August 8.  No $\simgreat 3\sigma$ infrared sources are visible at the location in the pipeline-processed mosaics.  Source photometry is challenging due to source crowding, particularly at longer wavelengths.  The object is located between two WINGS $JK$ objects within $\la5$~arcsec \citep{Valentinuzzi09}, with $JK$ magnitudes of 21.21 and 18.64.  From WFPC2 $F814W$ data, these neighbouring objects may be composite.  Using the IDL aperture photometry routine {\it aper}  \citep{IDLastro}, we find these neighbouring objects to have AB magnitudes of $15.3\pm0.1$ and $15.6\pm0.2$ at 4.5$\mu$m, and of $14.8\pm0.1$ and $16.1\pm0.2$ at 8.0$\mu$m.  Within $\sim1.4$~arcsec of WINGS J1348, we can set lower bounds of 18.9 at 4.5$\mu$m and 19.8 at 8.0$\mu$m.  At 4.5$\mu$m, the object may be very faintly visible, but it is impossible to distinguish from weak overlap of the neighbouring objects' PSFs.  MIPS shows an upper limit of 15.9 at 24$\mu$m (or 1.6 mJy), but is likely dominated by contributions from these bright neighbouring objects within the 5-arcsec extraction circle.  This confusion is worse at longer wavelengths, and so we ignore the lower-resolution 70$\mu$m and 160$\mu$m data.

WINGS J1348 was observed in the far infrared (FIR) by {\it Herschel} PACS\footnote{http://herschel.esac.esa.int} at 70$\mu$m and 160$\mu$m on 2010 January 21, and at 100$\mu$m, and 160$\mu$m on 2009 December 23 for 571s per visit.  Inspection finds no sources present within the 50 per cent encircled energy radius ($\sim3.5$-arcsec, 4-arcsec and 7-arcsec for 70$\mu$m, 100$\mu$m, 160$\mu$m respectively) at $\simgreat2\sigma$.  The {\it HerschelSpot} exposure time calculator shows that {\it Herschel} PACS is sensitive to point source fluxes $F_{FIR}(70\mu\rm{m})=15$~mJy, $F_{FIR}(100\mu\rm{m})=18$~mJy, and $F_{FIR}(160\mu\rm{m})=35$~mJy, to $\sim3\sigma$.

A broad band plot of these limits from UV to $160\mu\rm{m}$ will be presented in Section \ref{Archival}, in conjunction with a direct comparison to Seyfert 2 galaxies.

\subsubsection{Magellan Echellette Spectrograph}

On 2011 March 5, we obtained an 1800 s optical spectrum using the Magellan Echellette Spectrograph (MagE) on the Clay Telescope at Las Campanas Observatory, in order to determine the likelihood that WINGS
J1348 is a member of Abell 1795 as opposed to a line-of-sight coincidence.  MagE observed the galaxy over 15 spectral orders ranging
from 3645 \AA\ to 9465 \AA.  We used a 1-arcsec slit and had resolution $R=4100$.  The airmass at the target coordinates was 1.77 and seeing was 0.9~arcsec.

Despite the signal-to-noise ratio obtained for the continuum (S/N$\;\sim2$), a 30 minute spectrum should be sufficient to detect emission lines from a background AGN or
(for example) star-forming galaxy.  There are, however, no obvious emission lines that cannot also be attributed to instrumental
effects or poor sky subtraction.  For comparison, \cite{Xiao11} observe the emission-line AGN GH06 \citep{GH04}, $V\sim19$, $z=0.100$ for 1800 s and obtain continuum S/N$=10$.  

Due to the low signal-to-noise, further analysis of the MagE data is strongly interpretation-dependent.  We therefore defer such analysis to \S\ref{sec:member}, where we consider the MagE data in the context of cluster membership and available photometry.

\subsubsection{ROSAT and EUVE}


Observations of A1795 have been performed by {\it ROSAT} \citep{ROSAT} and the {\it Extreme Ultraviolet Explorer} \citep[{\it EUVE};][]{EUVE}, and we have investigated the archives of those missions to search for possible associations with the X-ray emission from WINGS~J1348.  We find no evidence for a pre-existing X-ray source in {\it ROSAT}.  We do, however, find strong evidence that this X-ray flare may be related to a bright $EUVE$ transient observed in 1998 by \cite{BBK99}.

The closest source to WINGS~J1348 in the Second Rosat PSPC Catalog \citep{2RXP} is the core of the A1795 ICM.  We therefore obtain the most stringent $ROSAT$ upper limit at the position of WINGS J1348 using archival $ROSAT$ HRI data from 1997 July 27.  We find a $\sim1\sigma$ upper limit of $F_X\la1.6\times 10^{-14}$~\es, a factor of 13 below the brightest \cha\ $F_X$.  To obtain this limit, we determine a count rate within an 80 per cent encircled energy radius ($\sim5.5$~arcsec) and convert to unabsorbed flux using WebPIMMS and the best-fitting model from CXO1.

$EUVE$\footnote{http://heasarc.gsfc.nasa.gov/docs/euve/euvegof.html} 
observed the extreme ultraviolet
(0.016--0.163~keV) with three scanning telescopes, as well as a fourth telescope capable of pointed spectroscopy in four bands and Deep Survey pointed observations.  In particular, the Deep Survey (DS) LexB filter was sensitive to photon energies approaching soft X-rays, reaching an effective area of $\sim25$ cm$^{-2}$ at $\sim0.14$ keV, with a half power bandwidth of $\sim0.063$ keV.  Pointed observations with $EUVE$ were therefore in principle moderately sensitive to TDFs in the brightest band of their spectrum.  Effective areas with other $EUVE$ instruments were at least a factor of 2 lower, and typically $\la1$ cm$^{-2}$ at peak sensitivity, requiring very long observation times and low $N_H$ column densities to gather meaningful data.

Examining archival $EUVE$ data, we find that A1795 (including WINGS J1348) was observed with Deep Survey pointings 7 times between 1997 January and 2000 July (summarized in Table \ref{EUVE}).  Inspection of the images reveals a bright transient near the core (Northwest) of A1795 during the 1998 March 27 observation, and fluxes within $\sim2\sigma$ of the background during the later $EUVE$ observations.  This bright transient was in fact reported by \cite{BBK99} as a startlingly unlikely contaminant to their observations of the A1795 ICM, and went essentially unexplored beyond this consideration.

\begin{table}
\small
\begin{center}
\caption{Transient count rates from EUVE observations of A1795\label{EUVE}}
\begin{tabular}{ccccr}
&&&&\\
\hline\hline
Date			&	Exposure	&	Counts		& $R_R$	&	$R_O$	\\
			&	(s)		&				& counts s$^{-1}$	& counts s$^{-1}$	\\
\hline
1997-02-03	&	90004	&	5273			& 0.0144	&	0.0000	 \\
1998-03-27	&	70787	&	7041			& 0.0476	&	0.0332	\\
1999-01-03	&	23334	&	2081			& 0.0189	&	0.0045	\\
1999-01-05	&	25028	&	2120			& 0.0202	&	0.0058	\\
1999-05-29	&	57559	&	4390			& 0.0143	&	$<0.0020$	\\
1999-05-31	&	14544	&	1140			& 0.0146	&	$<0.0020$\\
1999-07-07	&	73506	&	5921			& 0.0083	&	$<0.0020$\\
\hline
\end{tabular}
\end{center}
\begin{flushleft}
Counts are for total source and background within a 40 pixel radius of source peak brightness.  $R_R$ is the total background-subtracted rate for that region.  $R_O$ is the estimated rate of the flare, subject to $\sim0.002$~counts s$^{-1}$ source+background $2\sigma$ statistical error, or the upper limit for $R_R\sim R_O$.  The source region includes the entire bright diffuse emission region for A1795. \\ $R_O(\rm{1997-02-03})=0$ by definition, given $R_O(\rm{Date})=R_R(\rm{Date})-R_R(\rm{1997-02-03})$.
\end{flushleft}
\end{table}

\begin{figure*}
\includegraphics[angle=270,width=6in]{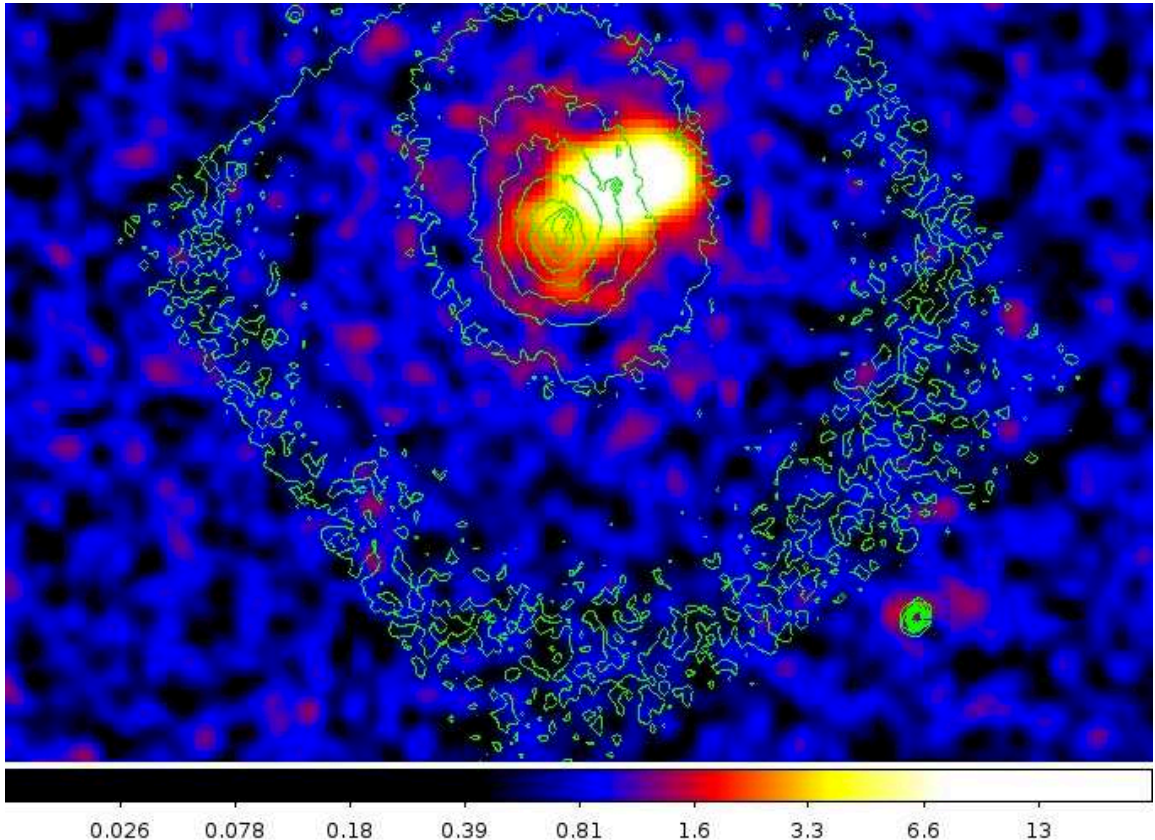}
\caption[{\it EUVE} image of A1795 Flare with \cha\ Contours]{{\it EUVE} image of A1795 flare with \cha\ (CXO1) contours.  The {\it EUVE} image is smoothed with a gaussian kernel of 2 pixel radius.  The colour scale corresponds to brightness in counts per square pixel.  Green \cha\ contours are lines of equal brightness on a square root scale.  Square
ACIS-S chip structure (8-arcmin~$\times$~8-arcmin) is evident in these contours.  The entire image is 12-arcmin~$\times$~9-arcmin.  The WINGS J1348 flare at peak luminosity is immediately ($\sim45$~arcsec) northwest of the bright centre of the diffuse ICM, which is more easily identified with \cha.  The \cha\ source is a
$\sim1$-arcsec contour structure coincident with the centre of the
{\it EUVE} source.  2E 1346+2646 is in the southwest corner of the image.  {\it EUVE} PSF structure is evident in the form of bright wings $\sim30$~arcsec to the west of both WINGS J1348 and 2E 1346+2646.}
\label{euve_cxc_overlay}
\end{figure*}

The positional location of the $EUVE$ transient is consistent with that
of WINGS J1348, and can be seen in Figure \ref{euve_cxc_overlay}.  The measured $EUVE$ PSF is
large ($\sim 24\;$arcsec FWHM) and therefore encompasses several cluster field
objects.  The pointing accuracy of $EUVE$ is small, however ($\sim2\;$arcsec) and
the core of the PSF is a few ($\sim4\;$arcsec) pixels wide \citep{Lewis93} with
relative centroiding accuracy of $\sim1\;$pixel \citep{Sirk97}.

In order to compute the $EUVE$ source positions, we correct the astrometry of the 1998
March 27 epoch using the bright nearby ($\sim6\;$arcmin) AGN 2E~1346+2646
\citep{Veron10}, visible also in CXO1, as well as the centre
of brightness contours overlaid to the diffuse ICM emission. For $EUVE$,
we assume the best position of a bright source to be the centroid of the
brightest PSF component.

The relative position we found for the $EUVE$ transient is within the
$\sim4$-arcsec $EUVE$ centroiding accuracy of WINGS~J1348.  The only bright
$WFPC2$ counterparts within the $EUVE$ FWHM which could plausibly contaminate
the $EUVE$ centroid are classified by SDSS as stars and are clearly
point-like in $WFPC2$: SDSS~J134850.01+263554.5 (here: A1795-S1) and SDSS~J134849.21+263550.5 (A1795-S2).  Unlike WINGS J1348, neither of these objects is a $\simgreat 2\sigma$ X-ray emitter in any of the \cha\ epochs.  A1795-S1 has $g\sim21.18$, and A1795-S2 is $\sim0.3$ magnitudes fainter in all $griz$ bands.  

The count rates in Table \ref{EUVE} are determined from the archival images as follows: using the \ciao\ tool $dmstat$, we extracted the total number of counts $N_T$ from a 40-pixel ($\sim2.7$-arcmin) circle centred on the peak of \cha\ transient source, with $EUVE$ coordinates corrected for astrometry relative to \cha\ observations.  This circle is large relative to the diffuse cluster emission, and encompasses both the cluster ICM and the TDFC to minimize uncertainty due to variations in the contribution from A1795, which is easily the brightest contributor to background.  Photon counts $N_B$ from instrumental and extended background outside of the cluster background are extracted by an annulus about the same circular region with an area $A_B$ 3 times  greater than the source area $A_S$.  As the transient and the A1795 ICM are easily the brightest $EUVE$ objects in this region, variations due to contributions from point sources are negligible.  The source region count rate is therefore $R_R=(N_T-N_B\times A_S/A_B)/t_E$, where $t_E$ is the observation exposure time.  Counts in the 1997 $EUVE$ epoch are assumed to originate entirely from the ICM, therefore we find the flare count rate $R_F=R_R(n)-R_R(1)$ where d is the number of the relevant $EUVE$ epoch, $n=1$ corresponding to the 1997 epoch.  We find the bright source is detected over the integrated emission of the entire cluster from 1998 March 27 to 1999 January 5, and place a limit of $R_F<4\times10^{-3}\;\rm{ s}^{-1}$ between 1999 May 29 and 1999 July 23.  Image inspection shows that the source remains resolved with varying levels of brightness  in all later images, however, and all source locations are consistent with the position of the \cha\ source to within a fraction of the PSF, $\sim20$~arcsec.  In principle an actual measurement of the count rate during these later epochs might be possible with careful selection of the source and background regions via detailed analysis of the $EUVE$ PSF, however here we only establish a relatively conservative upper limit relative to the A1795 ICM.

We have not attempted to correct the count rate for the dead spot in the Deep Survey detector, a $\sim1.7$-arcmin region with reduced response (as much as $\sim75$ per cent) due to the 1993 January observations of the bright EUV source HZ 43 \citep{EUVEdead,DrakePrivate10}.  Later observations circumvented this problem by aiming off-axis, and the accuracy to be gained by correcting for the dead spot appears small relative to the inherent uncertainties of our subsequent spectral modelling.

Although the position of the $EUVE$ flare is consistent with that of the \cha\ flare, the crowded cluster field and instrumental limitations of $EUVE$ create a significant challenge for a more confident association of these two sources.  Our primary scientific conclusions are independent of this association, however, and may be made without reference to the $EUVE$ data.  We therefore treat the $EUVE$ data with more detail in \S\ref{Archival}, in the context of WINGS~J1348 as a possible tidal flare host.



\section{Discussion}

\subsection{Observational and Theoretical Background}

In order to place our new results in context, we first provide brief
reviews of both previous observations and estimates of tidal flare
rates.

Although we expect tidal disruption flares to be among the most
luminous observable astrophysical events, with the total kinetic
energy of ejected debris exceeding that of supernovae at $10^{51}$ erg
or more \citep{Ulmer99}, these flares have thus far been challenging to observe.
While they should occur within AGNs and may contribute significantly
to the faint end of the AGN X-ray luminosity function \citep{Milos06}, they
will be easiest to identify in a quiescent galaxy, where they can be
distinguished from typical variable disc accretion in an AGN.

Existing theoretical studies predict disruption rates of 1 event per
$10^4 - 10^5$ years per galaxy, \citep[e.g.][]{MT99}, a rate that has been
supported by observational studies using ROSAT \citep{Donley02} . The
most optimistic predictions increase that rate by an order of
magnitude, assuming the rate is dominated by large numbers of dwarf spheroidal galaxies in clusters \citep{Wang04}.

Several candidate events have been observed \citep{Komossa05,
  Komossa04, Halpern04}, but some of the most convincing evidence of
tidal disruptions comes from \galex\ detections of UV flares with
optical and (sometimes) X-ray components as observed by
\cite{Gezari06,Gezari08,Gezari09,Gezari12}.  Ongoing studies using the XMM Slew
Survey \citep{Esquej07,Esquej08,Saxton12} also report TDFs as bright 
supersoft X-ray sources, but an
extensive study of the Chandra Deep Field \citep{Luo08} made no
detections, consistent with maximum rates comparable to $10^{-4}$ per
galaxy per year for $L \ga 10^{43}$ \es.  About 20
such candidate events have been identified to date (see the above
references), so the statistical conclusions that can be reached thus far
are highly tentative, especially given uncertainties as to the reality of any given tidal disruption flare candidate (TDFC).

The rate by which tidal flares occur should also act as an indicator
for the distribution of black holes in the galaxy population.  The
effect may be particularly pronounced according to the theoretical
calculations of \cite{Wang04} if a significant fraction of nucleated
dwarf spheroidal galaxies harbor MBHs at their centres.  Given that dwarf
spheroidals are a very numerous component of the galaxy distribution
\citep[see, for example][]{Jenkins07}, if lower mass MBHs flare more often than more massive MBHs, they may dominate the flare rate if they contribute at all.  But more recent work by
\cite{Merritt09} suggests lower mass MBHs may produce such flares more
rarely even if dwarf galaxies do possess MBHs.  As noted
in the introduction, determining the population of MBHs in dwarf
spheroids through such indicators as tidal disruption events will also affect
predicted rates of MBH-MBH mergers and extreme mass ratio inspirals
(EMRIs).  

\subsection{Derived Galaxy Properties}

\subsubsection{Cluster Membership}
\label{sec:member}


Establishing the distance to the host galaxy is critical to determining the properties of both the flare and the flare's host galaxy.  In this section we discuss different methods of estimating the distance to WINGS~J1348.\\

\noindent{\bf Line-of-Sight Probability:} While a spectrum of sufficient quality would allow a redshift-derived distance determination, in a sufficiently rich galaxy cluster, in lieu of such data we may make a probabilistic estimate of the galaxy's distance relative to the cluster.  The number of galaxies in the cluster relative to the number of galaxies in the field above a selected magnitude suggests that any given galaxy in the field of view may have a high probability of being a cluster member.  

Although two bands of photometric detection are inadequate for the purposes of a photometric redshift, we can begin to address the issue of the host galaxy's distance by other means.  At the most basic level, we can examine, as in \cite{Maksym10}, the probability that any flare of unknown host type is a cluster member based on its projected radius from the cluster core.  Comparing the approximate background number surface density of galaxies at the projected radius of WINGS J1348 ($\theta_p\sim2$~arcmin--3~arcmin) to the outskirts ($\theta_p\sim10$~arcmin--20~arcmin) where line-of-sight galaxies dominate, we find that $\sim27$ per cent of all galaxies to the limit of WINGS at the WINGS~J1348 angular separation of $\theta_p\sim2.5$~arcmin.  The inherent likelihood of cluster membership (absent all other considerations) is therefore high but inconclusive.\\

\begin{figure}
\centering$
\begin{array}{c}
\includegraphics[trim=0in 0in 0in 0.7in,clip,width=2.7in,angle=90]{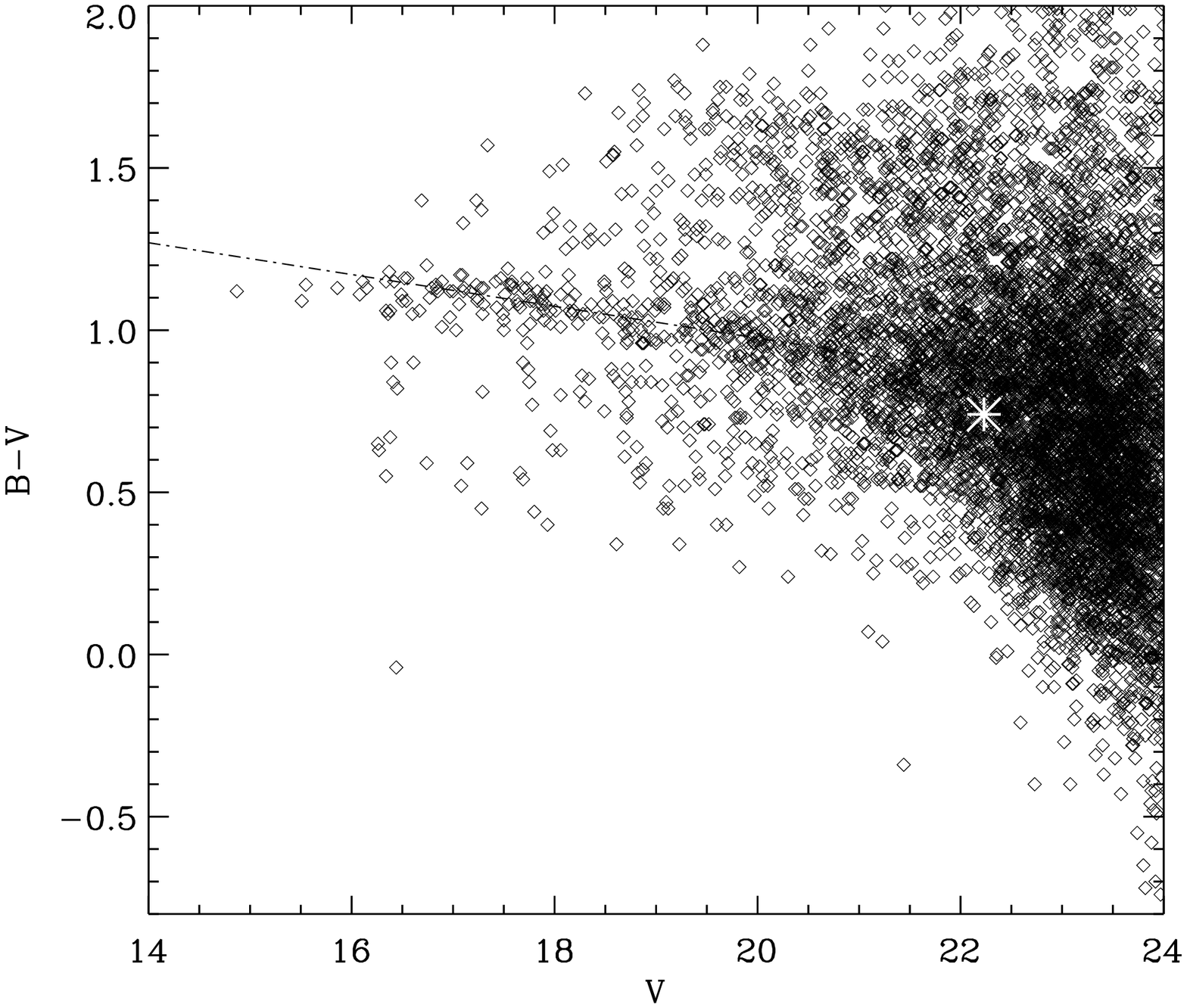} \\
\includegraphics[width=2.5in,angle=90]{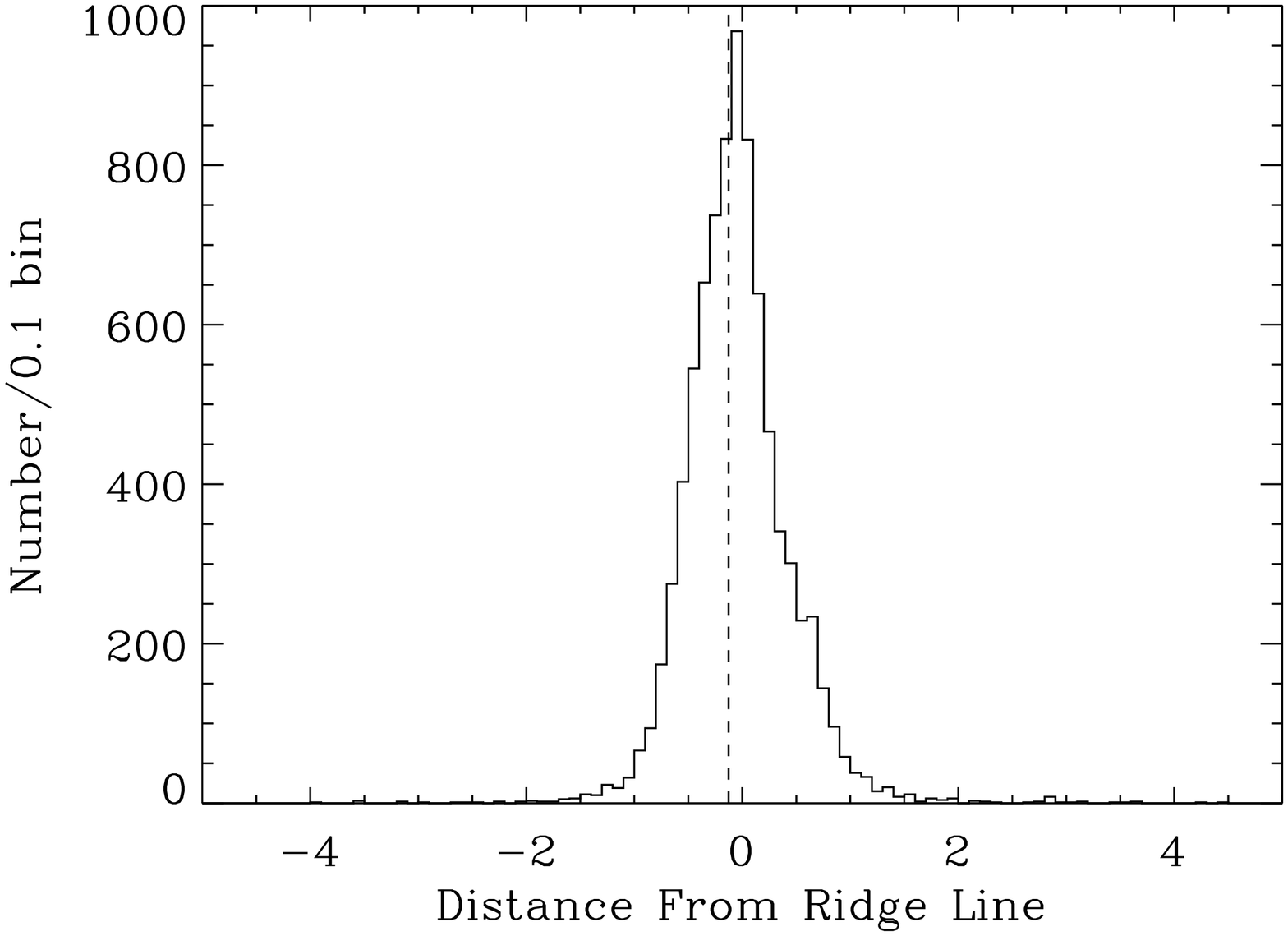}
\end{array}$
\caption[a) Colour-Magnitude diagram b) Cluster ridge line histogram]{Top: Colour-Magnitude diagram.  Black diamonds indicate A1795 galaxies in the WINGS survey \citep{Varela09}.  The dot-dashed line indicates the best-fitting ridge line for the cluster.  The white cross represents WINGS J1348, flare host galaxy, and is consistent with cluster membership.  Bottom: Histogram of distance from cluster ridge line, in magnitudes.   The dashed line indicates WINGS J1348, again illustrating consistency with cluster membership.}
\label{cmr}
\end{figure}

\noindent{\bf Photometric Constraints:} By plotting WINGS J1348 on a colour-magnitude diagram of cluster and background sources, as per \cite{LopezCruz04}, we can make a more substantive comparison.  If the host galaxy is a member of A1795, we expect it to fall on or near the cluster's `ridge line'.  In Fig. \ref{cmr} (top) we plot colour ($B-V$) vs magnitude $V$ using values from \cite{Varela09}.  As can readily be seen, WINGS J1348 is close to the best-fitting ridge line of cluster galaxies.  In Fig. \ref{cmr} (bottom), the histogram of cluster galaxies as a function of distance from the ridge line also shows that WINGS J1348 falls within one of the two bins closest to the ridge line, or within $1\sigma$.  This analysis supports the proposition that the host galaxy is a cluster member at $z\sim0.062$.  However, its faintness suggests that it may also lie in or near the locus formed predominantly by background galaxies at V$\simgreat22$.  A consequence of its colour on this diagram, as per \cite{Brusa07}, is that if WINGS J1348 is a bright background AGN, it should be at $z\simgreat3$.  This effect can be explained by the blue continuum of strong AGNs, which only inverts as Ly$\alpha$ shifts into the $V$-band.  

With no confident detection of the object via WINGS $JK$ or {\it Spitzer}, we have no evidence of, say, a bright rest-wavelength component at 4000\AA\ or 1000\AA, redshifted by $z\sim2$ or more.  We can also ask: would SDSS be expected to detect WINGS J1348 in redder bands ($r, i, z$) if the galaxy were a luminous background AGN or a faint dwarf galaxy in A1795.  Using the photometric conversions from \cite{Jordi06}, we expect WINGS J1348 to have a SDSS-equivalent $g\sim22.9$.  Although SDSS DR7 sources are typically only detectable $g\la22.2$ with 95 per cent reliability, the catalog extends to greater depth in some regions, and contains 67 spectroscopically confirmed QSOs at $z>2.7$ with $22.5<g<23.5$ and $\sigma_g< 0.15$.  All such QSOs have measured $i>20.5$ and $\sigma_i<0.05$.  These are significantly brighter than the average SDSS DR7 limit $i\sim21.3$ and the local sample, which reaches $i\sim22.1$ within 2~arcsec of WINGS J1348 with the requirement $\sigma_i<0.2$.  A cluster dwarf galaxy, however, might remain undetected with those limits.  \cite{Penny12} find dwarf ellipticals of comparable $V$ in the Perseus cluster to have $0.8<V-I<1.1$, which implies $I>21.4$ if it is applicable to A1795.  Or, using the \cite{Jordi06} photometric conversions and assuming $I>R>V$, we find $i>21.9$, which implies a significant fraction of the A1795 dwarf population is below the detection limits of SDSS, not just at $V\sim22.5$ but in redder bands as well.

If WINGS J1348 is a member of A1795, its redness ($B-V$) is more consistent with an early-type galaxy, i.e. Sa or S0 \citep{Fukugita95}, suggesting an older stellar population that might be expected of a dwarf spheroidal or ultra-compact dwarf galaxy \citep{Evsti08,THT09}, such as are common in the cores of galaxy clusters \citep[e.g.][]{Gregg09}.

Likewise, the galaxies in \cite{Dale07} typically have flux ratios of {\it Spitzer IRAC} $3.4\mu$m-to-$V$ of $F_{3.4\mu\rm{m}}/F_{V}\la1$, whereas the $z\sim1$ AGNs in \cite{Konidaris07} have $F_{3.4\mu\rm{m}}/F_{V}\simgreat70$.  By comparison, upper limits from the $IRAC$ observations of A1795 show $F_{3.4\mu\rm{m}}/F_{V}\la5$ for WINGS J1348.  This is again more consistent with WINGS J1348 being a dwarf cluster member than a background AGN.\\

\noindent{\bf Optical Spectrum:} 

To create a normalized uniform spectrum, we used a modified version of the
make\_1d\_echelle\_norm IDL script \citep{PR07}, which fits each spectral
order to a low-order normalized polynomial and returns the spectral intensity relative to that polynomial.  
The spectrum shows several large-scale features indicative of instrumental effects or poor sky
subtraction.  The spectral regions with the largest deviations from the norm, in particular, may indicate areas of order overlap
where sensitivity is poor for either order.  Continuum signal-to-noise is very low, reaching S/N$\;\sim 2$ at greatest sensitivity in the
4th order, near 6317 \AA.

The low signal-to-noise ratio severely complicates the identification of strong absorption lines typical to an early-type cluster
member (such as Ca H+K near rest wavelength 3950 \AA, \ion{Mg}{1} at 5174\AA, \ion{Na}{1} D at 5893\AA, and the Balmer series).  In the case of a
cluster member at $z\sim0.062$, the vast majority of these lines would be at points of the spectrum where instrumental effects are most
significant.  If WINGS J1348 is indeed a dwarf galaxy in A1795, we would not expect any prominent emission lines associated with the galaxy, and the galaxy would also likely have a low metallicity, resulting in inherently weak absorption features which are therefore also more difficult to identify.  The low surface brightness common to dwarf galaxies would also impair the ability to identify any spectral features. 
In principle, a longer observation taken with a more sensitive instrument would allow positive identification of such
lines, as was done in \cite{Maksym10}.  Such analysis is not possible, however, given the quality of the available spectrum.

We can, however, make an independent statistical test of our default position of cluster membership (given available photometric
evidence, as well as our previous spectral arguments against various background galaxies), which appears to at least support the
plausibility of cluster membership.  We tested this hypothesis by assembling the spectral orders into a single normalized log-binned spectrum and
cross-correlating the spectrum of WINGS J1348 with reliable spectra of known stars and galaxies.  Prior to cross-correlation, the spectra are resampled into logarithmically scaled bins appropriate to the scale of the instrument.  The presence of lag in the spectral
cross-correlation is indicative of the object's redshift. 

The noisiest parts of the spectrum are typically at the blue end, and we therefore found it advantageous to impose a blue end cut-off
in wavelength ($\lambda _C$), below which spectral bins were excluded from the cross-correlation.  However, the choice of $\lambda _C$ can be very subjective and
we therefore explored a range of $\lambda _C$ (3768.49--7072.93) in order to examine the cut-off effect on the cross-correlation function.  We likewise
explored a range of lags not only to search for cross-correlation (or the absence thereof) at the redshift of the galaxy, but also to compare
against signals due to instrumental effects, and to search for other possible cross-correlations at modest ($z\la 0.3$) redshifts.  We find a cross-correlation feature at $z\sim0.062$ that varies between $\sim2\sigma$ and $\sim10\sigma$ depending upon the choice of $\lambda _C$.  



We find a strong cross-correlation feature at $z\sim0.0$ for a wide range of $\lambda _C$, no matter
which objects are compared.  This feature demonstrates $\sim100$\AA\ scale or greater effects due to the failure of spectral calibration
and renormalization to eliminate all instrumental and sky features.  Our first important test of the validity of this technique is to
compare strong-signal MagE spectra from an early-type galaxy of known redshift ($z\sim0.0075$) and a stars (HR2049) of G
spectral type, expected to dominate an early-type galactic spectrum.  At $z\sim0.008$ (very close to the bright $z\sim0.0$ feature), we
see a faint cross-correlation line at $\sim12$ lag for even $\lambda _C$ as low as $\sim3767$\AA, and becoming more prominent for $\lambda
_C\simgreat$4600\AA.  We are also able to distinguish false cross-correlation features such as overlap between instrumental lines and the \ion{Ca}{2} triple.  
This analysis can be examined in greater detail in \cite{MaksymThesis}.


This trial analysis serves to demonstrate some of the limits and capabilities of this cross-correlation technique for the purposes of our analysis:
namely, that while we cannot unambiguously {\it prove} the expected redshift of an unknown galaxy given the presence of
instrument-dependent cross-correlation effects, we can {\it disprove} cluster membership of an expected spectral type by showing a
lack of cross-correlation between WINGS J1348 and an object with known absorption lines.  We also eliminate certain correlation-derived redshifts as instrumental effects.\\

\noindent{\bf Summary:}  While we cannot unambiguously determine the distance to the flare host galaxy, WINGS J1648, various circumstantial elements support the proposition that the galaxy is a member of A1795 at $z\sim0.062$, including the galaxy's position on a colour-magnitude diagram of available WINGS data for the cluster, lack of strong emission lines in the Magellan optical spectrum, and cross-correlation of the galaxy's Magellan spectrum with spectra from known objects.  We must therefore include this proposition in our analysis and consider it our default assumption, although we will entertain higher redshifts as probabilistically feasible.  A longer duration observation from a more sensitive instrument would allow more confident determination of the galaxy redshift and distance.

\subsubsection{\Mbh--$L_{Bulge}$ Black Hole Mass \label{MBH}}

Assuming the galaxy is member of A1795 at $z\sim0.062$, with an absolute magnitude of $M_V=-14.8$, the host must be a very low-mass dwarf galaxy \citep[for example,][predict a total stellar mass of $\sim1.3\times10^{8}\;$\Msun\ for WINGS~J1348 photometry]{Bell03}.  Such a host is sufficiently faint that we cannot be certain that the \Mbh--$\sigma$ and \Mbh--$L$ relations continue to hold at such low galaxy masses.  See, for example, \cite{Gultekin09b}, who use dynamically confirmed MBHs down to $M_{V,Bulge}\sim-16.42$.  The extrapolated \cite{Gultekin09b} \Mbh--$L$ relation predicts log(\Mbh/\Msun)$\sim5.4$ at most, in the case that the galaxy luminosity is dominated by its bulge (compare to \cite{Lauer07}, which predicts log(\Mbh/\Msun)$\sim4.9$).  These estimates are dominated by the uncertainty in the respective relations used.

\subsection{Was the Event a Tidal Flare?}

As in \cite{Maksym10}, we consider explanations of more common variable, X-ray luminous objects before addressing the event in terms of tidal disruption flares.  We find that the WINGS~J1348 TDFC is very similar in most respects to the flare in Abell 1689 \cite{Maksym10}, other recent observational examples \citep[e.g.][]{Esquej07,Esquej08,Lin11,Saxton12}, and basic theoretical predictions.  A tidal flare is therefore the best explanation for the event, even in the case that WINGS J1348 is a background galaxy rather than a cluster member.

\subsubsection{A Galactic Foreground Object?} 

As with Abell 1689, the high galactic latitude (77\degr) greatly reduces the probability of the flare arising from a line-of-sight foreground object.  From the luminosity profile in Fig. \ref{hstdata} and from the WINGS catalog \citep{Varela09}, we see that WINGS J1348 is an extended source and is therefore likely a galaxy.  As in \cite{Maksym10}, the object is unlikely to be a quiescent low-mass X-ray binary (qLMXB) due to its extremely soft spectrum and low $L_X$(0.5--2.5 keV) $\la10^{29}$ \es\ estimated at 1 kpc.  And at $B-V=0.8$, the faintness of the associated optical object rules out a flaring main sequence star or X-ray binary donor out to 40 Mpc.  A donor star with the same $V\sim22.5$ must be at $>10$ kpc for $M_V<15$.

\subsubsection{A Highly Variable AGN?}

As in \cite{Maksym10}, we consider the plausibility of AGNs as an explanation for the flare associated with WINGS J1348.  As in \cite{Maksym10}, this flare has X-ray emission that is significantly softer (with a photon power law of index $\Gamma\simgreat 4$) and more variable (50 times, vs. a few) than is typical for AGNs ($\Gamma\la 2.5$ and a factor of a few, respectively).  But similarly large variations are common for Narrow Line Seyfert 1 (NLS1) galaxies, and have been observed in other AGNs as attributed to changes in the absorption column density \citep{Risaliti05}, MBH binary-disc interactions \citep{LV96}, and other mechanisms that are well-summarized elsewhere \citep{Maksym10,Saxton12}.  As in these previous analysis, the presence of a persistent AGN does not in itself eliminate a TDE as an explanation for a dramatic X-ray or UV flare \citep[see, e.g., the flaring quasar Sharov 21, as per][]{Meusinger10}.  However, the ability to demonstrate a lack of evidence for persistent accretion-driven emission (as would be expected in a galaxy hosting an AGN) provides a strong argument against one of the most likely alternate explanations for this phenomenon.  

For the purpose of constraining a putative AGN explanation, the observed $F_X$(2.0--8.0 keV) allows us to derive a {\it redshift-dependent} ($z=0.062$) upper limit of $L_X\la5.5\times10^{40}$ \es\ on 1999 December 20, with later upper limits typically between $L_X\la8.4\times10^{40}$ \es\  and $L_X\la2.0\times10^{41}$ \es.  Note also that the limits we have established are for the integrated $F_X$ of the entire galaxy (which is unresolved in \cha) and have not modelled any possible contribution from an X-ray binary population.   These limits are consistent with normal galaxies, or with low-luminosity AGN (LLAGN) populations \citep[$L_X\la1.4\times10^{42}$ \es,][]{Ho08}, which also puts WINGS J1348 in a regime where previous observations have indicated that strong X-ray variability is unlikely on short \citep[$\la$~day scales,][]{Eracleous02} as well as longer time-scales: \cite{Young12} find that LLAGNs have suppressed variability on time-scales of months or years relative to the $L_X$-variance trend that they establish for AGNs with $L_X\simgreat10^{41}$ \es.  The limits we place on a quiescent AGN from the derived $L_X(0.2-8.0\;\rm{keV})$ are also comparable to those used by \cite{Gezari12} to  place limits on the existence of a pre-existing AGN at $z\sim0.17$.

Instances of AGNs with highly variable X-ray spectra that lack a hard ($\simgreat2$~keV) component are known.  The X-ray behaviour of WINGS~J1348  demonstrates several basic similarities in comparison to 2XMM J123103.2+110648 \citep[2XMM~J1231;][]{Terashima12}, a highly-variable AGN which also lacks significant 2--10~keV emission and has similar $L_X$ to WINGS~J1348 ($\simgreat2\times10^{42}$~\es).  The softer 0.2--1.0 HR of WINGS~J1348 (Fig. \ref{hardness}) in its low state could be evidence of spectral flattening at a higher accretion rate and hence Comptonisation characteristic of near-Eddington accretion, as per \cite{Terashima12}.  The power law slope of WINGS~J1348 is comparable to 2XMM~J1231, although WINGS~J1348 requires a cooler {\tt diskbb} ($kT\sim0.1$ vs. $kT\sim0.18$).  

But there are two major differences between WINGS~J1348 and 2XMM~J1231.  First, variability of WINGS~J1348 is large compared to 2XMM~J1231, a factor of $\simgreat50$ vs. $\sim3$ for 2XMM~J1231.  Unlike 2XMM~J1231, the extreme variability of  WINGS~J1348 appears to be truly transient, although constraints on the variability of 2XMM~J1231 are limited by only $\sim1.5$ years of useful X-ray observations.  Secondly, as will subsequently  be shown, 2XMM~J1231 demonstrates emission lines characteristic of AGNs, which should be detected in the Magellan spectrum if WINGS~J1348 hosted a similar low-mass AGN \citep{Ho12}.

{\it A Narrow-Line Seyfert 1 Galaxy or Similar Object? --} 
Although AGN X-ray variability by a factor of 50 times or more is rare, several instances have been observed, and have been summarized in \cite{Maksym10} and \cite{Saxton12}.  Such variability has been attributed in different instances to changes in absorbing column density, disc state transition, and interaction of a companion object with the accretion disc of the persistent AGN.  In all cases, however, optical emission lines characteristic of AGNs should be expected from the host galaxy if the central engine is not normally heavily obscured.  On a basic level, the very low $N_H$ column density and constraints on infrared emission are consistent with a dwarf galaxy rather than a dusty AGN.  The decreasing HR along with $F_X(0.2-2.0)$~keV, as per Fig. \ref{hardness}, is inconsistent with an increase in cold absorption and favors some combination of intrinsic softening and luminosity decay.

\begin{figure*}
\centering
\begin{subfigure}{3.25in}
\includegraphics[width=3.1in]{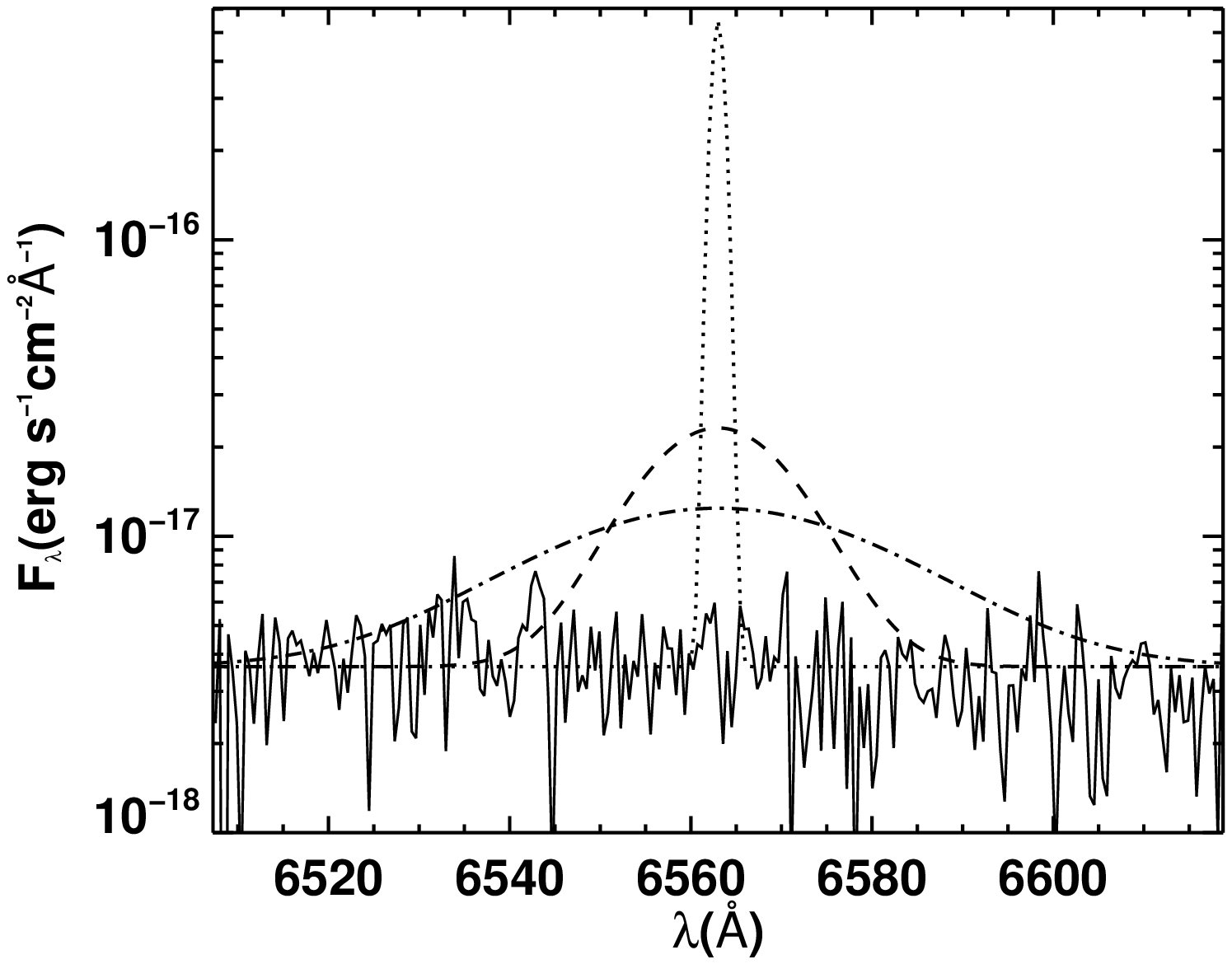}
\end{subfigure}
\qquad
\begin{subfigure}{3.25in}
\includegraphics[width=3.1in]{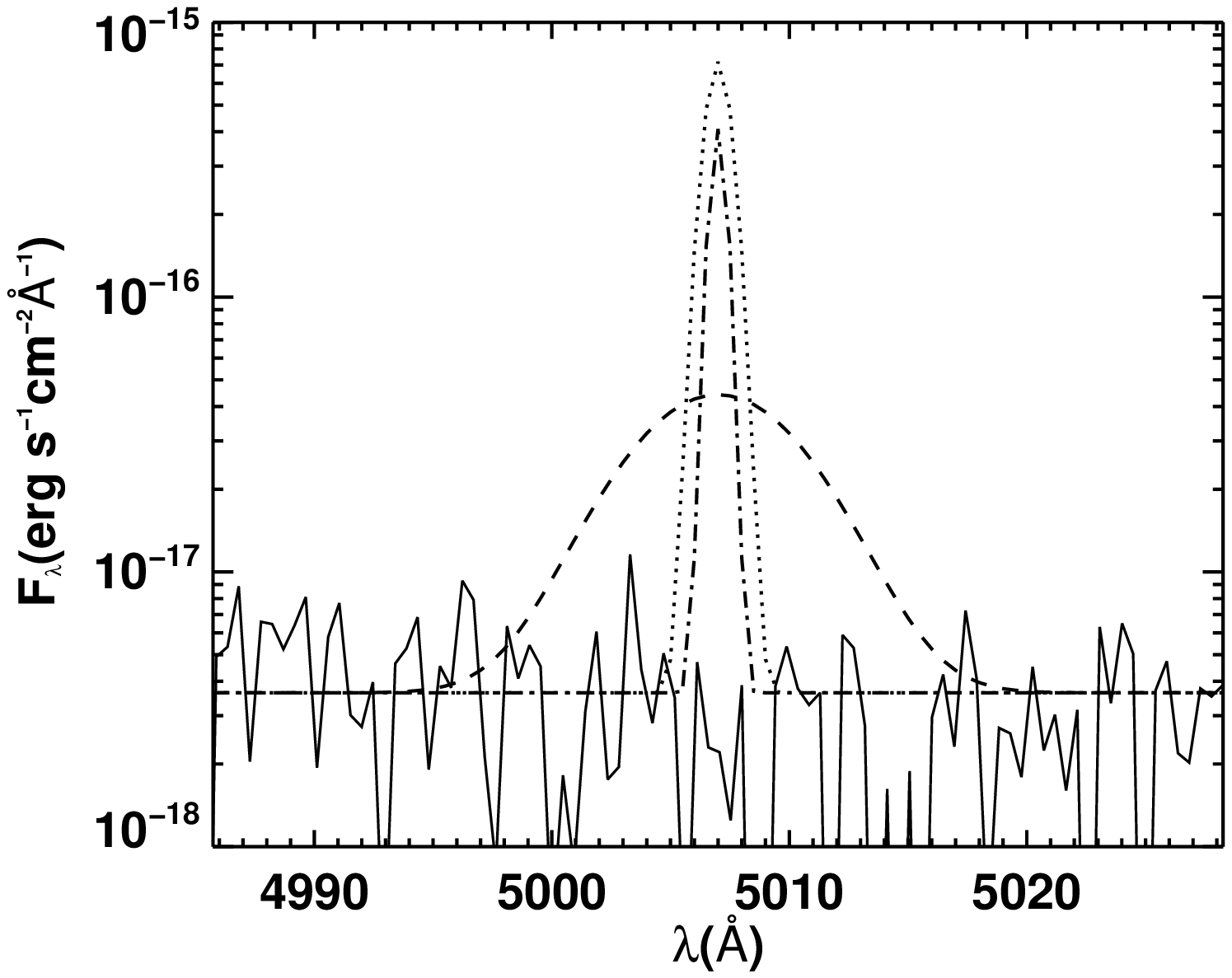}
\end{subfigure}
\caption[OIII and Halpha line profiles]{Simulated gaussian profiles (dashed lines) of typical AGN emission for H$\alpha$ (left) and [\ion{O}{3}] 
(right), overlaid with Magellan spectroscopy of WINGS~J1348, assumed to be at rest frame z=0.062.  Magellan data are normalized to V=22.5.  Dashed lines: line fluxes are derived from \cite{Panessa06}, assuming $F_X$(2--10)~keV equal to \cha\ upper limits.  Assumed FWHM are 900~km~s$^{-1}$ (H$\alpha$) and 500~km~s$^{-1}$ ([\ion{O}{3}]).  Dotted lines: $F$(H$\alpha,[\ion{O}{3}])=10^{-15}\;$\ecmss\ with FWHM(H$\alpha,[\ion{O}{3}])=50$\;km\;s$^{-1}$, comparable to the highly variable Seyfert galaxy in \cite{Terashima12} and \cite{Ho12} scaled to $z=0.062$.  Dot-dash lines: also as per \cite{Panessa06}, but in the extreme cases of FWHM(H$\alpha)=2000\;$km~s$^{-1}$ and FWHM$([\ion{O}{3}])=50\;$km~s$^{-1}$.
Data noise is typical for portions of the Magellan spectrum uncontaminated by atmospheric lines.}
\label{agnlines}
\end{figure*}

The NUV nondetection from {\it GALEX} implies that any Seyfert 1 type galaxy would have a baseline X-ray flux that is significantly lower than the upper limits we have established.  As per Table \ref{XrayHard}, CXO1 sets the lowest hard band upper limit, $F_X(2.0-8.0\;\rm{keV})\simless5.1\times10^{-14}\;$\ecmss.  We use the flux density at 2.0 keV from the best-fit hard band model to infer the ultraviolet ($\lambda=2500$~\AA) flux density using the range of values for the slope explored by \cite{GH07} such that $-2\simless\alpha_{\rm ox}\simless-1$, with $\alpha_{\rm ox}\equiv-0.3838\;\rm{log}\;(\mathnormal{f}_{\textup{2500\;\AA}}/\mathnormal{f}_{\textup{2 keV}})$ as per \cite{Strateva05}.  We would expect NUV$\simless22$ for all $\alpha_{\rm ox}$, significantly brighter than the {\it GALEX} limiting magnitude NUV$\sim23$ for all cases.

As in \cite{Maksym10}, we can address the issue of whether WINGS J1348 is an AGN at the redshift of the galaxy cluster from the limits on likely AGN emission lines, in particular  for those expected of NLS1s which can have the supersoft X-ray spectra and extreme variability typically expected of tidal flares.  At the expected $z=0.062$ wavelengths of H$\beta$ at 4861\AA, [\ion{O}{3}] at 5007\AA, and H$\alpha$ at 6563\AA, we fit a gaussian profile plus continuum for the amplitude of a 1$\sigma$ emission line undetectable in the noise of the Magellan spectrum.  We begin with assumed Balmer full width half maxima (FWHM) typical of NLS1s, 2000 km$\;\rm{s}^{-1}$, and [\ion{O}{3}] $\lambda$5007 has FWHM=500 km~s$^{-1}$.  Simulating a line with triple the amplitude, we confirmed that such a line would be detectable in the Magellan spectrum at $3\sigma$.  From these simulated lines, we infer 3$\sigma$ upper limits to the equivalent width (EW) of EW(${\rm H\beta})\la45$\AA, EW([\ion{O}{3}])$\;\la27$\AA, and EW(${\rm H\alpha})\la21$\AA.  Derivation of line flux limits from these EW values is only possible with a flux-calibrated continuum, however.  Assuming the WINGS \citep{Varela09} value of $V=22.46$, this translates to flux limits of $F{\rm (H\beta)}\la1.7\times10^{-16}$ \ecmss, $F$([\ion{O}{3}])$\la1.0\times10^{-16}$ \ecmss, and $F{\rm (H\alpha)}\la1.4\times10^{-16}$ \ecmss\ (for H$\alpha$, assuming a linear flux continuum between $V$ and $F814W=21.5$).  

The mean value from active galaxies in \cite{GH07}, however, implies a broad line component of only $\sim900\;\rm{km}\;\rm{s}^{-1}$ for FWHM$_{\rm{H}\alpha}$.  If the continuum for WINGS J1348 is noise-dominated, we can therefore rescale $\propto\sqrt{\rm{FWHM}}$ such that $F{\rm (H\alpha)}\la1.1\times10^{-16}$ \ecmss.  Note that FWHM$_{\rm{H}\alpha}$ and FWHM$_{\rm{H}\beta}$ are typically comparable, but $F({\rm{H}\alpha})/F({\rm{H}\beta})\sim3.5$, making H$\alpha$ the more obvious indicator of a putative underlying AGN.  

The narrow core of an hypothetical [\ion{O}{3}] line is observed to correlate with \Mbh\ and would have a smaller equivalent width, as low as $\sim50\;\rm{km}\;{s}^{-1}$ for $\Mbh\sim2\times10^5\;\Msun$ \citep{Xiao11}.  Such a low-mass black hole would therefore have more stringent flux limits, down to $F$([\ion{O}{3}])$\la3.2\times10^{-17}$ \ecmss.  For comparison, 2XMM~J1231 would have $F([\ion{O}{3}]$, H$\alpha$)\;$\sim10^{-15}$~\ecmss at $z\sim0.062$ \citep{Ho12}.

\cite{Panessa06} examine $L_X\rm{(2.0-10.0\;keV)}$ vs. $L{\rm (H\alpha)}$ and $L$([\ion{O}{3}]) in a local sample of Seyfert galaxies, and the results from their total Seyfert population show that an AGN emitting in X-rays just below our $L_X{(2.0-8.0\;\rm{keV})}$ upper limits would have $L{\rm (H\alpha)}=3.5\times10^{39}\;$\es\ and $L$([\ion{O}{3}])$\;=3.2\times10^{39}\;$\es\ at $z=0.062$.  This corresponds to 
$F{\rm (H\alpha)}\la4.1\times10^{-16}$ \ecmss\ and  $F$([\ion{O}{3}])$\la3.6\times10^{-16}$ \ecmss, which is a factor of a few above our detection thresholds in each case.  The spectroscopic limits on an underlying cluster member AGN are therefore stringent relative to the X-ray limits, despite the noisy continuum, and imply any AGN must be quite intrinsically weak.  To illustrate, we have simulated gaussian profiles for $F{\rm (H\alpha)}$ and $F$([\ion{O}{3}])  derived from \citep{Panessa06} in Fig. \ref{agnlines}.  Since we do not measure these lines, we consider lines of varying hypothetical widths between extreme broad (2000 km$\;\rm{s}^{-1}$) and narrow (50 km$\;\rm{s}^{-1}$) cases, as well as lines comparable to those in \cite{Ho12}.  In all cases, such hypothetical lines would be detected.

If we consider only NLS1s and assume an X-ray power law of $\Gamma\sim4.21$ and the high-state flux of $F_X\sim2.08\times10^{-13}$ \ecmss, the expected line fluxes of a NLS1 at $z\sim0.062$ would be F(${\rm H\beta})=4.2\times10^{-15}$ \ecmss, and F([\ion{O}{3}])$=2.08\times10^{-15}$ \ecmss, based on the 90 per cent of sample values from \cite{Grupeetal04}.  Again, this analysis is similar to \cite{Maksym10}.  The low-state upper limits to $F_X$ are more stringent, but still a factor of a few through 2004, depending upon the spectral model.  All spectral models produce $\Gamma\simgreat4$, however, which is markedly softer than examples from \cite{Grupe04} of comparable $L_X$  ($\Gamma\la3$).  

Note that  \cite{Panessa06} derive the line strengths of $F{\rm (H\alpha)}$ and $F$([\ion{O}{3}]) from galaxies with typical narrow-line region sizes of $\sim1000$\;pc, as in \cite{Bennert06}.  If WINGS~J1348 is in A1795, the small size of the galaxy might also imply a small narrow-line region, and therefore weaker $F{\rm (H\alpha)}$ and $F$([\ion{O}{3}]).  The results of \cite{Ho12}, however, imply that comparable narrow-line strengths to those of \cite{Panessa06} are possible for an active galaxy with an effective radius of only $\sim700\;$pc (see Fig. \ref{agnlines}).

With this caveat, we can confidently conclude that WINGS J1348 is not a NLS1 in A1795, and if the line flux limits derived for a galaxy at $z\sim0.062$ are typical for this spectrum, then at higher redshifts a NLS1 or other AGN might be excluded as well.  For example, \cite{Hainline11} find a mean EW(Ly$\alpha$)$=66.39\pm11.65$ and $\Delta v=197\pm10$ km~s$^{-1}$ in rest frame of their sample of UV-selected AGNs at $z\sim2-3$.  If the $B-V$ colour is due to redshifting of the Lyman break to $z\simgreat3$, this implies we have observed dramatic variability in a bright quasar with $L_X\simgreat2\times10^{46}$ \es\ and $M_V\la-24.5$ (yet atypically of an AGN, there is no sign of a $\Gamma\sim2.0$ X-ray power law).  But at $z\simgreat3$, Ly$\alpha$ will have redshifted to $\simgreat4800$\AA, with F(${\rm Ly\alpha}$) at least 5 times greater than the upper limits we have estimated.

\begin{figure}
\includegraphics[width=3.15in]{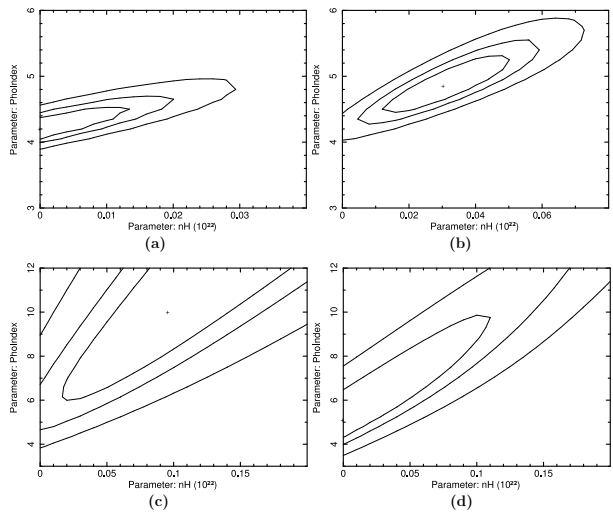}
\caption[Contour plots of confidence regions varying photon index $\Gamma$ and column density $N_H$ in $10^{22}\;\rm{cm}^{-2}$ for the first four epochs in Table \ref{XrayFits}]{Contour plots of confidence regions varying photon index $\Gamma$ and column density $N_H$  in $10^{22}\;\rm{cm}^{-2}$ for the first four epochs in Table \ref{XrayFits}.  Confidence levels from inner to outer contours for each subplot are as follows: a) \cha\, 1999--12--10, $\chi^2=$127,129,134.  b) \cha\, 2000--03--21, $\chi^2=$136, 138, 143.  c) \xmm\, 2000--03--21, $\chi^2=$165, 167, 170.  d) \cha\, 2002--06--10, $\chi^2=$174, 176, 181.}
\label{zpow_contour}
\end{figure}

As in \cite{Saxton12}, with high-quality X-ray data we can test the possible but somewhat contrived hypothesis that the luminosity evolution could be caused by a change in the column density in a persistent but normally obscured AGN, such as might be possible from the temporary opening of a hole or window in the absorption during the epochs of greatest observed luminosity.  As can be seen in Table \ref{XrayFits}, we modelled later epochs against the parameters of the X-ray spectral fit from the first and brightest epoch while varying only $N_H$.  We reject an explanation for X-ray variability based purely on evolution of the column density to $>99$ per cent confidence.  As in \cite{Saxton12}, a column density change approaching $\Delta N_H=10^{23}$ cm$^{-2}$ is required for a neutral absorber.  A more complicated evolution resulting from some combination of varying $\Gamma$ and $N_H$ is possible, as can be seen in Table \ref{XrayFits} and contour plots from the first four X-ray epochs in Fig. \ref{zpow_contour}.  However, there is significant degeneracy between $N_H$ and $\Gamma$, making it difficult to constrain these values independently in later, fainter observations.  As the observed $F_X$ declines, larger $N_H$ becomes permissible by assuming a softer spectrum.  All epochs are, however, consistent with modest evolution in $N_H$ and $\Gamma$. 

{\it A BL Lac Object. --}  As in \cite{Maksym10}, we examine the FIRST \citep{FIRST97} radio catalog and find no evidence of persistent radio emission that would be expected from a BL Lac Object or other AGN with persistent jet emission.  In addition, the X-ray spectrum ($\Gamma\simgreat4$) is markedly softer than that of typical BL Lacs \citep{Donato01}.

\subsubsection{Other Extragalactic Line-of-Sight Objects?}

As the A1689 flare \citep{Maksym10}, the flare in A1795 is poorly described as a supernova as compared to previous observations of X-ray luminous supernovae during their early evolution.  Although shock breakout models have been used to describe comparably luminous soft X-ray emission from supernovae \citep{Campana06,Soderberg08,GezariSN08}, such emission has been highly variable and of short duration ($\sim$~hours) as compared to the sustained emission of this flare over years.  More luminous emission, particularly from type IIn supernovae as shocked ejecta propagate into the interstellar medium at late times ($\sim$~months into the shell expansion phase), has been observed to approach $10^{42}$~\es, but while X-ray luminous supernovae may have a soft thermal component, these most luminous supernovae are typically described by high temperatures and hard spectra ($\simgreat8$ keV) \citep{Immler03,SP06,Immler08,DG12,PChandra12}.

Long-term, highly variable X-ray emission is also commonly seen amongst GRBs.  While recent evidence suggests that at least some GRBs are best described as resulting from a jet formed as the result of a TDE \citep{Bloom11,Zauderer11,Cenko12Swift,Burrows11}, most GRBs are thought to form by jet power from other means, such as via compact binary mergers or core-collapse supernovae \citep[e.g.][]{Berger11}.  In any event, X-ray emission from GRBs of any kind is typically described by $\Gamma\la2$, which is much harder than the observed thermal spectrum.  We therefore find that the X-ray observations cannot be construed to describe a jet-dominated phase of any kind.

At $L_X(\rm{0.2-2.0\;keV})\sim2\times10^{42}$ \es, the flare is also relatively faint compared to known X-ray selected TDFs, with a range of $\sim10^{42}-10^{44}$ \es\ described for classical supersoft examples \citep{Gezari09}.  This complicates luminosity-based arguments against a ULX as an explanation, as ULXs have also been known to reach $\sim10^{42}$ \es\ in their most extreme cases, such as ESO 243-49 HLX-1 \citep{Farrell09}.  HLX-1, however, is an unusual case and may in fact be a case of repeated accretion from a donor star by an intermediate mass black hole (IMBH) rather than sustained accretion by a stellar mass black hole or less massive object as is probably the case for less luminous ULXs \citep{Lasota11}.  Continued X-ray observations of both objects can be expected to clarify any differences between them.  In any event, the sustained low levels of emission from the flaring object in A1795 after 2004 are not consistent with continued cyclical emission as is observed in HLX-1.

We should also note that all comparison of the A1795 flare with competing extragalactic explanations that reach $\sim10^{42}$ \es\ are dependent upon the assumption that the host galaxy is a cluster member of A1795 (as we expect).  If the galaxy is in fact not within A1795, the low redshift of A1795 and the faintness of the host galaxy imply that the object would then be more distant, in which case the actual peak luminosity of the flare may become significantly greater than is typical for ULXs or supernovae.  But even in the most likely case of a galaxy within A1795, the most plausible non-TDF explanations all are less likely than a TDF.


\subsubsection{Tidal Flare Explanation}

On the basis of the preceding discussion, we will now proceed to discuss the event on the basis that it was a TDF.

The observed data are strongly consistent with the now well-established criteria laid forth in numerous theoretical \citep[e.g.][]{Rees88,Ulmer99,LR11} predictions and observational candidates \citep[][]{BKD96,KG99,KB99,Maksym10,Cappelluti09,Esquej08,Esquej07,Lin11,Saxton12} for a classical (i.e. not dominated by a beamed jet) flare from the tidal disruption of a star by an MBH.  Namely, we have identified a luminous ($L_X\simgreat2\times10^{42}$ \es), supersoft ($kT\sim0.09$, $\Gamma\sim4.21$) X-ray flare that is significantly above (in this case $\times50$) the quiescent $L_X$, consistent with the galactic nucleus, and poorly described by more common sources of luminous X-ray flares, such as may be explained by a persistent AGN, X-ray luminous supernova, bright galactic X-ray source, or other such phenomenon.  

Such a flare should be broadly consistent with the $t^{-5/3}$ decay expected to scale with Keplerian evolution of the debris accretion rate \citep{Rees88,Phinney89}, although recent theoretical work demonstrates numerous likely deviations from this picture.  Firstly, the early transition between the initial luminosity rise and $t^{-5/3}$ decay \citep{Lodato09} should vary depending on the density profile of the disrupted star.  There may also be significant deviations from $t^{-5/3}$ decay at later ($\simgreat10-1000$ days) time-scales as the fraction of material reaching the black hole evolves differently than the accretion rate during the initial super-Eddington phase, as well as exponential decay in X-rays several years post-disruption \citep{LR11}.  The X-ray light curve may also be significantly affected by temperature evolution of the accreting material and obscuration by ejecta and super-Eddington winds \citep{SQ09,SQ11}.    One possible explanation for the apparent spectral softening in Fig.  \ref{hardness} could be cooling of the disc as the accretion rate declines, as per \citep{SQ09}.

If not an artifact of \cha\ calibration, the soft excess in early X-ray spectra (CXO1) suggests a bolometric correction $f_{bol}\gg1$ may be necessary, as compared to the more modest values of $f_{bol}$ used by \cite{LNM02} and \cite{Maksym10}.  We could easily find $f_{bol}\simgreat10$ if the X-ray emission may be reasonably approximated by some sort of two-component model (such as for hot emission near $R_{ISCO}$ or the Comptonisation-hardened emission of a hot disc \citep[e.g.][]{ST93,ST95a,ST95b,LNM02}, combined with cooler emission near \Rd\ or from an expanding shell of ejecta, as detailed in \cite{Saxton12}.  This is also consistent with the \xmm\ observation, which is well-fit to $kT\sim0.025$ keV, but for which a harder, fainter blackbody component might be lost in the diffuse ICM emission due to the large PSF of \xmm.  As is evident from our fits of CXO1 to {\tt compbb}, a pure Comptonised blackbody is difficult to constrain without reliable FUV or U,B photometry during the first 2--3 epochs, and tends to yield results which are probably unphysical.

Previous UV observations \citep{Gezari09,Gezari12} have hinted at a strong discrepancy between models of emission for UV-selected TDFs as compared to X-ray selected TDFs.  Although numerous flares have been detected in both regimes, UV-selected tidal flares commonly have weak-to-nonexistent X-ray detections, suggesting these observations may probe different regimes (such as the inner and outer disc, or disc accretion and diffuse envelope or wind) or be sensitive to different combinations of \Mbh\ and post-disruption evolutionary phase \citep{Gezari09,SQ09,SQ11,LR11}.  

High-quality X-ray data for non-relativistic TDFs have been rare post-ROSAT, with some very recent exceptions \citep{Lin11,Saxton12}, and the early epochs of the WINGS J1348 flare allow us to examine the problem of extrapolating between the UV and X-ray regimes in some detail.  The soft excess in epoch CXO1 may be indicative of the Wien tail of a cooler, more extended component that is more easily seen in UV.  This component is barely detected at $z\sim0.062$ during an early epoch when the ACIS-S sensitivity is near the mission maximum sensitivity.  Indeed, the fact that we have observed a TDF at such low redshift using early observations of a \cha\ target serves to illustrate the limits of \cha\ for follow-up of TDFs selected by other means, as with PS1-10jh \citep{Gezari12}.  For if the WINGS J1348 flare were observed at the same redshift as PS1-10jh for the same (10 ks) duration as with PS1-10jh and with the current effective quantum efficiency of ACIS-S \citep{ACIScontam}, it would only produce $\sim9\pm4$ counts.  This also implies that the A1689 flare observed in \cite{Maksym10} may have a much larger $f_{bol}$ than was previously determined, as it would have been impossible to identify a similar soft excess using such late observations with ACIS-I at $z\sim0.19$, if the TDF in A1795 is physically similar to the one in A1689.

\subsection{Comparison of Observations to Tidal Flare Models}

\subsubsection{Light Curve Decay}

\begin{figure}
\includegraphics[angle=0,width=3.15in]{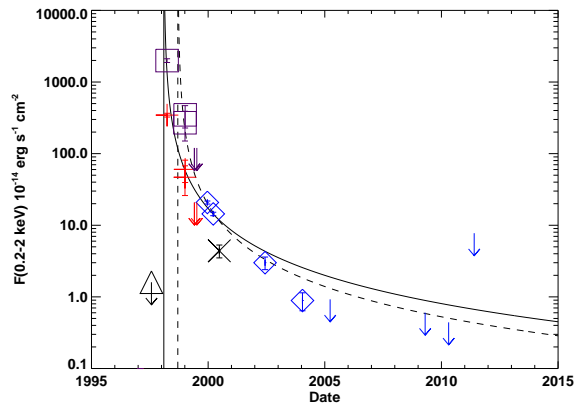}
\caption[Model-dependent X-ray flux evolution for WINGS~J1348]{Model-dependent X-ray flux evolution for WINGS~J1348: At the distance of Abell 1795 ($z\sim0.062$), $10^{-14}$~\ecmss$\sim10^{41}$ \es. \cha\ (blue diamonds) and \xmm\ (black $\times$) fluxes are assumed $\Gamma=4.21$ as per Table \ref{XrayFits}.  Arrows are upper limits.  Purple boxes and red crosses represent upper and lower estimates of $F_X(\rm{0.2-2.0\;keV})$  corresponding to models of {\it EUVE} data as described in the text.  The {\it ROSAT} upper limit is $\sim1.6\times10^{-14}$~\es\ at Date~$=1997.57$ (triangle and arrow).  The dashed line describes $t^{-5/3}$ decay for $t_0-t_D\sim500$~days, where $t_0$ is the earliest \cha\ data point.  The solid line represents $t^{-5/3}$ for $t_0-t_D\sim50$~days where $t_0$ is the earliest {\it EUVE} data point.  Vertical lines are placed at $t_0$ for their respective curves.}
\label{flux_lc}
\end{figure}

Assuming the WINGS J1348 flare was due to a TDF, it has one of the best-sampled X-ray light curves among TDFs reported to date, exceeded only by the jet-dominated flares Swift J1644 \citep{Bloom11,Burrows11} and Swift J2058 \citep{Cenko12Swift}, as well as the non-relativistic TDF in SDSS J1201 \citep{Saxton12}.  Unlike these previous examples, WINGS J1348 has been identified via archival analysis of repeated observations of a \cha\ calibration target rather than an active monitoring campaign.  As a result, timely follow-up observations have not been possible.  WINGS J1348 has, however, been monitored over a much longer period of time than any of these previous flares, over a span of $\sim12$ years covering \cha\ and \xmm\ observations alone.

Although, as mentioned in the previous section, we expect significant deviations from idealized $t^{-5/3}$ decay, we find this model to be an excellent fit to the evolution of the \cha\ count rate in Fig. \ref{counts_lc} (with the \xmm\ epoch converted to a \cha\ rate using WebPIMMS\footnote{http://heasarc.nasa.gov/Tools/w3pimms.html} for the best-fitting spectral model).  The \cha\ count rate has, however, varied over time due to contaminant buildup on the detector \cite{ACIScontam}.  And as we see in Table \ref{XrayFits}, the estimated flux may be heavily model-dependent.  For the sake of simplicity, we therefore compare $t^{-5/3}$ decay to $L_X$ determined uniformly from a power law at the best fit for CXO1, $\Gamma=4.21$, and absorbed by galactic $N_H$ (Fig. \ref{flux_lc}, including data described in subsequent subsections).  Fig. \ref{flux_lc} includes data from all epochs, including \cha, \xmm, and {\it EUVE}.  \cha\ fluxes are plotted in blue, and the sole \xmm\ flux is a black `$\times$'.

Although count rates for {\it EUVE} are high in early epochs, the lack of energy resolution in its Deep Survey camera requires any flux estimates to be extrapolated from models fit to later data at other (i.e. X-ray) wavelengths.  We therefore plot two plausible scalings of the X-ray light curve according to the early {\it EUVE} data points.  Red symbols are lower estimates, and assume $F_X(\rm{0.2-2.0~keV})$ in the final (upper limit) {\it EUVE} epochs is comparable to CXO1 and directly proportional to the {\it EUVE} count rate.  Purple symbols assume the count rates of the same {\it EUVE} epochs correspond to a modestly absorbed ($\la3$ times galactic, $N_H=3.0\times10^{20}$~cm$^{-2}$) $\Gamma=4.21$ power law.  The corresponding $F_X(\rm{0.2-2.0~keV})$ from the model is plotted.

The disruption time $t_D$ is difficult to constrain from the light curve given the large inherent uncertainties in the $L_{bol}$ \citep[including uncertainty in $N_H$, the soft spectrum, the strong model dependency of $L_{bol}$, and the potential for the spectral shape to evolve with time, as per][]{LR11}, and the latest upper limit we can place using $ROSAT$ extends to $\sim2$ years prior to the earliest \cha\ observation.  As per \cite{Burrows11} and \cite{Saxton12}, there may also be significant short-term variability with respect to an expected $t^{-5/3}$ evolution of $L_X$.  The jet-dominated TDF Swift J1644 has been highly variable on short time-scales $\la$~days in X-rays, and the very well-sampled instance of a more classical flare in \cite{Saxton12} also provides evidence of significant variability and deviations from $t^{-5/3}$ decay on short time-scales.  Without {\it EUVE}, the \cha\ and \xmm\ light curve evolution are consistent with $t_0-t_D\sim500$.  Evolution of $t^{-5/3}$ for $t_0-t_D\sim500$~days is plotted in Fig. \ref{flux_lc} as a dashed line, where $t_0$ is the earliest \cha\ data point.  This line describes the absorbed power-law spectrum {\it EUVE} model well except for the earliest and brightest epoch.  The solid line represents $t^{-5/3}$ for $t_0-t_D\sim50$~days where $t_0$ is the earliest {\it EUVE} data point.  In this case, the solid line is well-described by the simple linear relationship between {\it EUVE} count rate and X-ray flux, except at late times where $F_X$ falls below $t^{-5/3}$ decay.  This divergence may be evidence for late ($\sim$~years post-disruption) band-specific exponential decay, as per \cite{LR11}. 

With regards to evolution of the X-ray spectra, we note that a possible increase in column density is consistent with spectral models (Table \ref{XrayFits}, Fig. \ref{zpow_contour}), but is insufficient to explain the evolution of $L_X$ entirely.  As \cite{Saxton12} note with respect to the likely TDF in SDSS J1201, the radiation-driven ejecta postulated by \cite{SQ09,SQ11} may create a time-dependent neutral absorber.  Unlike SDSS J1201, we do not have sufficient ultraviolet constraints to exclude such an effect.  However, $t_D$ is so early relative to our observations that we are likely well past the $\sim130$ day window, within which such effects are significant.

\subsubsection{Archival Extrapolation to Early Times\label{Archival}}

\noindent{\bf Is an Association Between the EUVE and WINGS J1348 Flares Likely?} The presence of this $EUVE$ emission is a challenge to interpret due to the limited resolution and sensitivity of the telescope.  The emission could, however, significantly influence on our understanding of the A1795 X-ray flare if it has the same origin as the \cha\ flare.  We therefore attempt to assess the likelihood of coincident origin in some detail.

In a crowded field such as A1795, the greatest difficulty is determining whether the $EUVE$ flare is associated with a given optical source without knowing its character {\it a priori}.  But bright $EUVE$ sources are quite rare, particularly extragalactic sources.  The $EUVE$ Faint Source Catalog \citep{Lampton97} lists 534 objects identified jointly with $ROSAT$ in the all-sky surveys of those instruments.  The odds of a source coincidence within $\sim1$~arcmin of a randomly chosen target are therefore $\sim1$ per $10^5$, or $\sim1$ source per $10^3$ for a randomly targeted \cha\ ACIS field.  And the evolution of the two flares are strongly consistent with each other, as follows.

The weak sensitivity of $EUVE$ and strong attenuation of EUV or soft X-rays near $\sim0.14$ keV by intervening material imply that such a bright source must either be of galactic origin, or an exceedingly bright extragalactic source at relatively low redshift \citep{DrakePrivate10}.  This also implies that any object observed by $EUVE$ should be easily visible by \cha\ or \xmm\ unless it is a true non-recurring transient \citep{DrakePrivate10}.  As the $EUVE$ flare is exceedingly bright in 1998 March 27, brighter than the entire A1795 core at $\sim0.14$ keV, we must therefore assume the source either to be of galactic origin, or be as X-ray luminous as a bright AGN (indeed in this case much brighter than nearby NLS1 2E 1346+2646, if only temporarily).  The $EUVE$ source remains bright at $\simgreat 3\sigma$ on a time-scale of $\sim1$ year, implying a gradual evolution in brightness.  The count rate declines by a factor of 6 in 0.77 years after its peak, and if later (1999 March) positive identifications are equivalent to (at a minimum) a $\sim2\sigma$ fluctuation in the  $\sim15$-arcsec core of the PSF, this is consistent with an additional decay by a factor of 6 over the next $\sim0.4$ years.  This implies a very gradual decay.  Extrapolating this trend to the first \cha\ epoch (6 times decay in $0.46$ years, conservatively assuming the light curve does not flatten as a power law would), we use the PIMMS flux conversion tool from $ciao$ version 3.4, and find that for a blackbody of $kT=0.025$ (neglecting any harder component, as none is seen in CXO1) and $N_H=3\times10^{20}\;\rm{cm}^{-2}$, $F_X\rm{(0.2-2.0)}\sim1\times10^{-14}$ \ecmss, while for a power law $\Gamma=4.21$ and $N_H=3\times10^{20}\;\rm{cm}^{-2}$, $F_X\rm{(0.2-2.0)}\sim8\times10^{-14}$ \ecmss.  Either of these fluxes would be easily detected in CXO1.  The lack of any other such source within $\sim15$~arcsec of CXO1 implies that CXO1 is most likely has the same origin.

We can also address the question from the other side, whether the persistence of the bright \cha\ source to earlier epochs is likely.  If the \cha\ source were a bright flare from a persistent AGN, there is no particular reason to believe that it must begin suddenly, in a span of months prior to the earliest epoch in a flare that is visible for $\sim4$~years.  Indeed, such extreme supersoft X-ray flares from AGNs are sufficiently rare in the literature that the coincidence with an extreme but unassociated $EUVE$ flare in a marginally softer ($\Delta E\sim 0.05-0.1$ keV) X-ray band would in itself be remarkable.  

A TDF explanation arises very naturally from the $EUVE$ emission, however, if the emission is associated with WINGS~J1348 (and with interesting consequences for the TDFC itself).  Later $EUVE$ detections are consistent with $t_D$ derived from \cha\ and \xmm\ data fit to a $t^{-5/3}$ decay, though the 1998 March $EUVE$ epoch is somewhat earlier.  As can be seen from Figure \ref{flux_lc}, however, a $t^{-5/3}$ can be easily fit to the new {\it EUVE} epoch, assuming a model that consistently scales the upper limit to $EUVE$ count rate in 1999 July to the flux modelled to CXO1.  To explore the plausibility of such models, we examine count rates and fluxes predicted by models similar to those considered for CXO1.  Previously, we found $F_X(0.2-2.0~\rm{keV})\sim9.0\times10^{-15}\times(R_F/10^{-4}\;\rm{counts~s}^{-1})$~\ecmss\ for a blackbody with $kT_{BB}=0.025$~keV (possibly appropriate for the soft excess component to CXO1), and $F_X(0.2-2.0~\rm{keV})\sim6.0\times10^{-14}\times(R_F/10^{-4}\;\rm{counts~s}^{-1})$~\ecmss\ for a power law with $\Gamma=4.21$ (a model well-fit to CXO1).  These models predict $L_X(0.2-2.0~\rm{keV})\sim3\times10^{43}$~\es\ and $L_X(0.2-2.0~\rm{keV})\sim2\times10^{44}$~\es\ respectively for a member of A1795 at $z\sim0.062$ at the peak of $EUVE$ emission.  If $f_{bol}\simgreat10$, as for a supersoft blackbody, the associated $L_{bol}$ are comparable to those predicted in the literature by numerous sources, including predictions of super-Eddington luminosities that make peak $L_X$ relatively insensitive to declining \Mbh\ ($\la10^6~\Msun$) \citep[see, for example][for a variety of predicted monochromatic luminosities at early stages]{LR11}.

Neutral absorption becomes a more significant factor at EUV energies compared to even the softest \cha\ energies, complicating spectral modelling without additional constraints at higher or lower energies, or even modest energy resolution.  As a result, all associated analysis depends upon order-of-magnitude estimates to provide meaningful constraints on potentially radically different physical scenarios.  

The uniform scaling factor representing an assumed linear relationship between {\it EUVE} count rate and  $F_X(0.2-2.0~\rm{keV})$, as assumed in Figure \ref{flux_lc}, is therefore a simple but reasonable approach, and consistent with a variety of plausible models.  For example, the $kT_{BB}\sim0.025$~keV model is consistent with only modest $L_X$ evolution between the latest {\it EUVE} epochs and the two-blackbody model of CXO1 that roughly approximates two regimes of emission, such as between \Rd\ and $R_{ISCO}$, and provides one possible explanation for the difficulty involved in fitting CXO1 to a single blackbody.  

This model requires a large $f_{bol}\sim25$ and suggests that if WINGS J1348 is a cluster member and the \cite{Lauer07} \Mbh$-\sigma$ relation scales to lower masses, $L_{bol}$ may be significantly super-Eddington (by a factor of 2) even at late times ($t-t_D\simgreat1.7$ years), possibly dominated by super-Eddington winds.  Sustained super-Eddington accretion also raises the question of whether a jet may have formed \citep{KP12, DeColle12}.  Alternately, if $L_{bol}\la L_{Edd}$, \Mbh\ must be an order of magnitude greater than the host bulge luminosity would imply.  

In any event, a large $f_{bol}$ complicates measurements of X-ray-selected TDFs using \cha, particularly more distant objects using ACIS-I \citep[such as in][]{Maksym10}, where very limited photon counts do not permit detailed spectral modelling.  In particular, estimates of \Mbh\ via $L_{Edd}$, $R_{ISCO}$, and $t_D$ may all be affected.\\

\noindent{\bf EUVE-Related Constraints on the WINGS J1348 Transient:} If, in any case, the $EUVE$ source is indeed associated with the WINGS J1348 flare, there are significant consequences for models of both WINGS J1348 and the associated flare.  If the flare is not from a galactic object, it is also highly unlikely to be from a persistent AGN, and cannot be one at high redshift.  To be less bright than the most luminous known quasars \citep[$\sim10^{46}$~\es, e.g.][]{Levan11}, we would find $D_L/D_{A1795}\la10(f_{bol}/5)^{-1/2}$, where $D_L=10D_{A1795}$ at $z\sim0.5$, even with a very modest $f_{bol}$ and $N_H$.  By comparison, $z<0.4$ for all $EUVE$ AGNs identified in \cite{Polomski97}.

This $EUVE$ constraint allows us to consider in greater detail alternate models of variability from background AGNs at $z\la0.4$.
  Variability in $L_X(0.2-2.0~\rm{keV})$ by a factor of 600, as for the more conservative $kT_{BB}=0.025$~keV model above, is exceedingly rare in AGNs.  By comparison, one of the most extremely variable known examples is the dimming of the supersoft NLS1 WPVS007 by a factor of 400 between 1990 and 1993 \citep{Grupe95a}.  But while NLS1s are capable of such dramatic variability, the host galaxy clearly cannot be a NLS1 with the given redshift constraint.  NLS1s \citep[e.g. as in][]{Zhou06} typically have flat, blue spectra, incompatible with $B-V=0.8$ at $z\la0.5$.  This would even be true of a heavily outflow-absorbed NLS1 such as WPVS001, which has a very blue spectrum at $\lambda>2500\;\AA$ \citep[rest frame,][]{Leighly09}.
  
A greater challenge to this sort of SED analysis would be to exclude the possibility of a flaring Seyfert 2 galaxy.  For example, the unabsorbed Seyfert 2 GSN 069 demonstrated variability of a factor of 200 over 10 years, which was attributed to a disc state transition by \cite{Saxton11AGN}.  Or a galaxy with strong absorption could hide a Seyfert 2 AGN which experiences an outburst.  \cite{Saxton11AGN} found that Seyfert 2 galaxies are actually more the most likely galaxy to exhibit strong long-term X-ray variability.  And Seyfert 2 galaxies may be intrinsically red compared to Seyfert 1 galaxies.  If $z\la0.5$, however, we find the photometric nondetections from SDSS $riz$ bands and the $Herschel PACS$ to be broadly incompatible with typical Seyfert 2 galaxies as follows.

To test a Seyfert 2 model for WINGS J1348, we examine all spectroscopically classified Seyfert 2 galaxies in the \cite{Veron10} catalog at $0.062<z<0.5$, with $23>V>22$ to find near-analogues of WINGS J1348 under available constraints from $EUVE$, \cha, and WINGS.  The catalog contains 50 such galaxies, with typical $z\sim0.4$.  All of these Seyfert 2 galaxies have $i<20.4$, compared to the limit of $i\sim21.3$ which we previously established near WINGS J1348 (see Sec. \ref{WSphot}), which is consistent with the typical Seyfert 2 properties \citep[see, for example, the composite Seyfert 2 spectrum used for the $Spitzer$ Extragalactic Performance Estimation Tool\footnote{http://ssc.spitzer.caltech.edu/warmmission/propkit /pet/expet/help.html}, as per][]{Schmitt97}.

\begin{figure}
\includegraphics[angle=0,width=3.15in,clip=true,trim="0in 0in 0in 0.19in"]{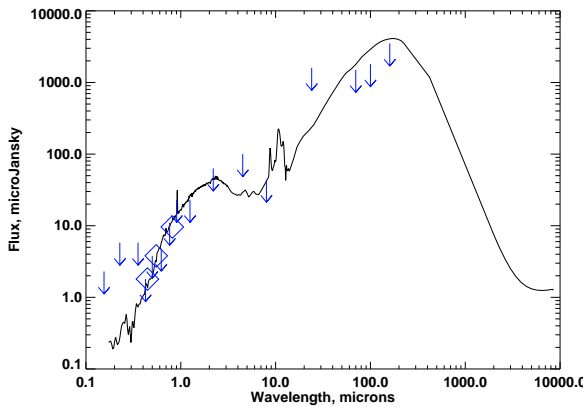}
\caption[Broadband SED of WINGS~J1348]{Broadband SED of WINGS~J1348, compared with a mean Seyfert 2 template from \cite{SWIRE} redshifted to $z\sim0.4$ and rescaled to match $B,V$ photometry.  Note that all data points except $B,V,F814W$ (diamonds) are upper limits.}
\label{sy2sed}
\end{figure}

Furthermore, an upper limit of $z\la0.5$ allows us to apply limits established via {\it Herschel} PACS, in order to constrain the presence of any Seyfert 2 features.  Seyfert 2 galaxies commonly display a prominent rest-frame FIR thermal component \citep[e.g.][]{RamosAlmeida11}, commonly attributed to some combination of warm dust (as for an AGN torus) and cool dust in star-forming regions.  The limits from {\it Spitzer} MIPS and {\it Herschel} PACS are sufficient to completely exclude a luminous infrared Seyfert 2 galaxy like Mrk 273  \citep{Brauher08} or Mrk 938 \citep{Esquej12} for $z\simgreat0.5$, whereas a fainter FIR object such as the nucleus of NGC 3081 \citep{RamosAlmeida11} might be permitted by these observations for  $z\simgreat0.1$.  In Figure \ref{sy2sed}, we compare the broadband photometry for WINGS~J1348 to a composite Seyfert 2 spectral energy distribution with mean values from \cite{SWIRE}, redshifted to $z=0.4$ and scaled to match the B,V photometric detections of WINGS~J1348.  Note that all other data points are upper limits, and are either at or slightly below the expected values from this template.

We therefore find broadband photometric constraints to be incompatible with a NLS1 interpretation, and to exclude a luminous infrared Seyfert 2 galaxy as the source of the flare.  LIR-faint Seyfert 2 galaxies are not excluded by virtue of the {\it Spitzer} and {\it Herschel} observations, but are incompatible with an SDSS nondetection at longer wavelengths.

Finally, at $L_X\simgreat10^{43}$~\es, we can confidently exclude both a ultraluminous X-ray source (ULX) of any sort or an X-ray luminous supernova \citep{Heil2009,Immler03,SP06,Immler08,DG12,PChandra12}.\\


\noindent{\bf Alternate Interpretations of the EUVE Flare:} While our previous analysis indicates a likely association between the $EUVE$ flare and the X-ray flare in WINGS 1348, we must consider the possibility that the $EUVE$ flare has arisen by coincidence from some other source within the field.  Within 15~arcsec of WINGS J1348 (the extent of the {\it EUVE} PSF), we find two SDSS photometric stars ($V<21$; SDSS J134849.21+263550.5, SDSS J134850.01+263554.5) and two very faint ($V>24$) objects undetected by SDSS and classified photometrically by WINGS as galaxies (WINGS J1348850.51+263558.5, WINGS J134849.64+264544.4).  

{\it EUV} flares have been known to occur as a result of outbursts from convective stellar atmospheres \citep{Audard00} or accretion on to a compact object, as is common for X-ray binaries \citep{Osten00}.  Both SDSS objects have similar colours $\pm0.1$, with $u>23$ and $\Delta\sim0.3$ magnitude difference from each other for all $griz$.  With $g-r\sim0.9$, $r-i\sim0.4$, these are well-described by the \cite{Lenz98} models for cool ($T\sim4000-4500$ K), low-metallicity main-sequence or giant K-stars.  Photometric parallaxes as per \cite{Juric08} place these objects each at $\sim4.2\pm0.4$~kpc, well above the galactic plane.  Although K-stars have been known to produce luminous flares, they are less common than for cooler M-dwarfs.  And the most luminous known flares of this type have had maximum $L_X(0.25-11.0~\rm{keV})\sim2\times10^{31}$~\es, with peak $F_X$ near 1 keV.  With a predicted $F_X(0.25-11.0~\rm{keV})\sim10^{-14}$~\ecmss\ at 4.2 kpc, the equivalent observed {\it EUVE} count rate is far too high to come from a stellar flare at that distance.  Under a range of power laws ($\Gamma=2.5-3.6$) discussed for {\it EUVE} stellar flares by \cite{Audard00}, we find the expected associated $L_X(0.25-11.0~\rm{keV})$ $\times100-1000$ greater than has been observed from even these extremely luminous stellar flares.

EUV flares with associated $L_X\rm{(0.2-2.0\;keV)}\sim10^{34}-10^{35}$ \es\ may be caused by an accreting compact object in a binary pair, such as a cataclysmic variable (CV, such as a  K-star - white dwarf pair) or low-mass X-ray binary (LMXB).  In particular, CVs may produce novae at intervals of decades or more.  Nova Cygni 1992 produced $EUVE$ count rates of $\sim0.11\;\rm{ct\; s}^{-1}$ at $\sim1.4\;$kpc with $N_H\sim3\times10^{21}\;\rm{cm}^{-2}$ \citep{SB94}.  At $\sim4.2\;$kpc and $N_H\sim10^{20}\;\rm{cm}^{-2}$, comparable luminosities are therefore attainable for a putative nova from these SDSS objects, although only $\sim25\;\rm{yr}^{-1}$ are typically expected in the Milky Way \citep{Matteucci03}, most of which can be expected from the galactic plane or globular clusters.  We have previously addressed the low probability of a chance coincidence for any {\it EUVE} source.  Post-flare X-ray flux limits imply that any such compact binary would be emitting at $L_X(0.2-2.0\;\rm{keV}, 2.0-8.0\;\rm{keV})\simless10^{31}\;$\es\  in quiescence.  But a compact companion is unlikely in any case, given optical photometry from both SDSS objects fails multiple colour criteria used by SEGUE \citep{SEGUE09} to describe binaries composed of a main sequence star and compact object.  An absence of emission lines in optical spectroscopy of these neighboring starlike objects would support this assessment.

Finally we consider whether the {\it EUVE} source may arise independently from one of the two WINGS faint non-stellar sources, despite the remarkable coincidence previously addressed.  Such faint galaxies may, like WINGS J1348, be either a dwarf galaxy in A1795 or a background source.  In either case the set of plausible interpretations remains similar, namely some large change in accretion by a MBH such as via a TDE or similarly large short-term variation in an AGN that is normally several orders of magnitude fainter in soft X-rays.  The only additional consideration is that (due to the lack of {\it EUVE} DS energy resolution) we must also consider the possibility of a hard transient such as a typical GRB with fast X-ray decline \citep[see][for a comparison]{Burrows11}.    In such a case, we would expect associated gamma ray emission detectable by all-sky monitors and consistent with the time constraints from the {\it EUVE} observations (discussed in more detail subsequently).\\



\noindent{\bf Comparison with Archival Gamma Ray Bursts:} Given the recent identification of Swift J1644 and Swift J2058, GRBs well-explained by beamed emission along the line of sight from a jet, which in turn was formed and sustained via a TDE \citep{Levan11,Bloom11,Burrows11,Zauderer11,KP12}, it is worth considering whether the flare from WINGS J1348 may be of a similar class, particularly given the possibility of sustained super-Eddington accretion, as above \citep{KP12,DeColle12}.  Such events may be as few as $10^{-6}$ of the total TDF population \citep{Bloom11,Cenko12Swift}, but wide-field hard X-ray monitors typical for GRB science missions have the potential for detection of such rare events early in their evolution, tightly constraining $t_D$ beyond what is possible for sporadic pointed observations and offering the opportunity to explore numerous other aspects of the flare.  The supersoft ($\Gamma\sim4.21$) spectrum argues against beamed emission, however CXO1 could occur significantly past the time at which total jet luminosity declines below thermal luminosity due to disc accretion \citep{KP12}.  Given the lack of energy resolution for the {\it EUVE} transient, however, we must also consider the possibility of a chance association with a more typical core-collapse or compact inspiral GRB.

To constrain any emission from a putative TDE jet, as well as to investigate the possibility that the associated {\it EUVE} source may be associated with a GRB of other origin, we examine archival records of the {\it Compton Gamma Ray Observatory} ({\it CGRO}).  In particular, \cite{Stern01} compiled a catalog of sources from the Burst and Transient Source Experiment on {\it CGRO} \citep{Fishman92}.  The BATSE Large Area Detector was sensitive to photons of $E_\gamma\sim30-1900$~keV, and capable of simultaneous monitoring of the entire sky between 1991 April and 2000 June.  The \cite{Stern01} catalog covers all triggered GRBs, as well as additional sources a limit of 0.1 photon~$\rm{s}^{-1}\rm{cm}^{-2}$ in $F_X(50-300~\rm{keV})$.  As the peak $\nu L_\nu$ for Swift J1644 should be roughly constant between $\sim4-4000$~keV \citep{Burrows11}, and the sensitivity of BATSE is comparable to the {\it Swift} Burst Alert Telescope at these energies \citep{Band06}, a comparable event should be sufficient to trigger a bright, $\sim$year-long event even at $z\sim0.35$ (vs $z\sim0.062$ for A1795).

The \cite{Stern01} catalog contains 25 sources between the date of the first {\it EUVE} epoch and the first \cha\ epoch where the positional displacement from WINGS J1348 is less than the positional uncertainty.  These uncertainties are quite large ($\simgreat 10\degr$ in most cases), such that it is difficult to say with any great confidence that a given GRB can or cannot be positively associated with the WINGS J1348 flare in any way.  The simplest explanation would therefore be that they are unassociated.  For the {\it EUVE} transient to be associated with a given GRB remains a remarkable coincidence given the large uncertainty in measurement for BATSE (10 GRBs between the 1997--02--03 and 1998--03--27 epochs, with error radius $\simgreat14\degr$, or $\sim1$ GRB per $10^3$ \cha\ fields).  We also note that no source appears in the BATSE Earth Occultation Catalog at this position \citep{Harmon04}, which is sensitive to $\sim1.1\times10^{-9}$~\ecmss\ from persistent 20~keV-1~MeV sources, or $10^{46}$~\es\ at $z\sim0.062$, 100 times less than the luminosity observed from Swift J1644.  

\subsection{The Massive Black Hole in WINGS J1348:}

As per \cite{Maksym10}, we have inferred \Mbh\ from the observed characteristics of the TDFC.  From the host bulge luminosity and \cite{Lauer07}, we expect $\Mbh(L)\sim 2.5\times10^5\Msun$.  As a first lower limit, we use the Eddington limit inferred from $F_X$ during CXO1 and find $\Mbh(L_{Edd,X})\simgreat1.3\times10^4~\Msun$.  Eddington-based constraints become more challenging, however, when we attempt to base them on early {\it EUVE} data or to estimate the bolometric correction $f_{bol}$.  The light curves we derive from {\it EUVE} data imply $L_X(\rm{0.2-2.0\;keV})\simgreat 2\times10^{43}$~\es\ for the earliest epochs, or $\Mbh(L_{Edd,X})\simgreat1.5\times10^5\;\Msun$.  This estimate remains compatible with $\Mbh-$Bulge from \cite{Lauer07}, as determined in \S\ref{MBH}.  However, if $f_{bol}\sim10-20$ (as previously discussed) then \Mbh\ could easily be an order of magnitude greater than expected.  Fits of CXO1 to {\tt diskbb} also imply $\Mbh\sim10^5\Msun$ (to the extent that simple multi-colour disk models are even applicable to the high-energy continuum of TDFs, which is uncertain).  Suppose alternately that {\tt ezdiskbb} may be used with a color ratio $f\sim3$, as in \cite{LNM02}.  In this case, $\Mbh\sim10^6\Msun$, in accordance with the larger estimate.  In any event, however, the extreme softness of the X-ray spectra suggests near-Eddington accretion.





\noindent{\bf Implications for the Host Galaxy} Although a TDF in a background galaxy ($z\simgreat0.07$) cannot be excluded, the implications for the host galaxy are quite interesting if WINGS J1348 is a member of A1795, as is most likely.  The discovery of a TDF in a $M_V\sim-14.7$ galaxy is an interesting opportunity to examine in detail a MBH identified to high confidence in a very small ($\sim300$ pc at $\sim1.195$ kpc/arcsec angular distance scale from the WFPC2 images) dwarf galaxy.

For comparison, Henize 2-10 hosts a startlingly large (log~$[\Mbh/\Msun]\sim6.3$) black hole for its relatively low stellar mass \citep[$\Mbh\sim3.7\times10^{9} $~\Msun within a 1 kpc core,][]{Reines11}, or $M_V\sim-18.8$ (derived from NED).  At $M_V\sim-14.4$, WINGS J1348 would be more than an order of magnitude less massive, assuming comparable mass-to-light ratios, and has 1/3 the spatial extent.  \cite{Reines11} indicate Henize 2-10 is already a challenge to explain from conventional MBH evolution models, so it would be interesting to determine \Mbh\ to high confidence for WINGS~J1348.

Even the existence of an MBH in such a tiny galaxy is interesting, as several more massive galaxies appear to lack conclusive evidence for an MBH compatible with the $\Mbh-\sigma$ relation \citep{GebhardtEtAl01,MFJ01,ValluriEtAl05}.    The inferred galactic mass $M_{gal}$ is low even compared to the bulge dynamical masses in the sample used by \cite{Jiang11}.  \cite{Strigari08} infer a central dark matter density of $\sim 0.1\;\rm{\Msun pc}^{-3}$ from dwarf satellite galaxies of the Milky way, which implies a stellar mass of 10$^6\;$\Msun\ in the central 300~pc of WINGS J1348.  The ratio of core stellar mass to \Mbh\ would therefore be at least 1:50 and possibly greater than 1:1, depending upon peak $L_{bol}$.

As WINGS J1348 appears quite close to the massive brightest cluster galaxy at the centre of A1795, 4C 26.42 ($\sim50$~kpc projected distance), a natural explanation for this strange juxtaposition would be a previous encounter with another galaxy.  Such encounters are common in the inner regions of clusters relative to field galaxies, and are thought to significantly drive cluster galaxy evolution and the diffuse intracluster light at optical wavelengths \citep{Moore96}.  An intriguing explanation might therefore be that some fraction of the stars in WINGS J1348 has been stripped via a previous tidal encounter.

The recoil from an uneven MBH-MBH merger could produce a fast-moving runaway MBH \citep{Bekenstein73} surrounded by a small cloud of gravitationally bound stars and possibly an elevated TDE rate \citep{KM08}.  Several such recoiling MBHs have been proposed in the literature \citep[see][for a review]{Komossa12}.  

As per \cite{MSK09} WINGS J1348 is sufficiently small to fit the description of a recoiling \Mbh, however only if we assume the recoiling MBH is large ($\Mbh\simgreat10^9$~\Msun) relative to the maximum radius allowable to disrupt a main sequence star without it falling directly into the event horizon, \citep[$\sim10^8$~\Msun; see, however,][for TDEs with $\Mbh\sim10^9$~\Msun\ via MBHs spinning under the Kerr metric]{Kesden12}.  

If WINGS J1348 is a galaxy in the background of A1795, however, the increased distance strengthens luminosity arguments against other more tenuous but theoretically possible explanations such as X-ray supernovae or ULXs of any sort.  Such a scenario therefore increases the likelihood that we are indeed describing a TDF.


\subsection{Tidal Disruption Rate from Abell 1795}

In \cite{Maksym10}, we determined a rate of tidal disruption $\gamma$ as a function of the number of TDFs observed in the rich galaxy cluster Abell 1689 over the course of 7 years of \cha\ observations and the population of galaxies expected to be present in the area subtended by those observations ($\sim1$ ACIS-I field, or $\sim16$-arcmin~$\times$~16-arcmin).  In principle we could attempt to apply similar analysis to A1795, although the observational conditions are radically different.  We have monitored $\sim200$ galaxies with \cha\ and \xmm\ over 13 years in the $\sim0.3$~Mpc core of A1795 with very deep photometry, compared to $\sim2000$ galaxies in the inner $\sim1.5$~Mpc of A1689, as determined via integration of its Schechter function.  

As a very rough approximation the combined rate would imply a disruption rate $\sim70$ per cent higher than previously determined.  However, a more thorough analysis would also include the outskirts of A1795 which are observed irregularly by \cha\ due in large part to the shape and size of its field-of-view, as well as 8 additional clusters in which we have found no flares that can be construed as originating from a TDE.  A revised estimate would therefore be closer to the lower estimates of \cite{Donley02}, \cite{Esquej08}, and \cite{Komossa12talk}, on the order of a few~$\times10^{-5}$~galaxy$^{-1}$~year$^{-1}$.  More detailed analysis is beyond the scope of this work, and will be examined in greater detail by a subsequent paper.  

\section{Conclusions}

In the course of our continuing program of galaxy cluster analysis, we have identified a luminous ($L_X\simgreat10^{42}$~\es) X-ray flare in the direction of Abell 1795.  The flare's supersoft X-ray spectrum ($\Gamma\simgreat4$), extreme variability (factor of $\simgreat40$ decrease in X-ray epochs alone), and long-duration transient nature consistent with $t^{-5/3}$ evolution of the accretion rate make it a strong candidate for a tidal disruption flare even without the consideration of {\it EUVE} data, which we argue is best explained as an earlier phase of the flare's evolution.  If so, then the total variability from the {\it EUVE} peak is at least a factor of 400 and may be as much as 40000, depending upon spectral modeling and associated bolometric corrections.  

The late-time observations with \cha\ and \xmm\ alone make this one of the best-sampled non-relativistic (non-jetted, non-beamed) tidal flares to date with modern high energy resolution X-ray detectors.  These observations span nearly $\sim12$ years of monitoring at intervals varying from days to years, thanks to a pre-existing \cha\ calibration program, and $>13$ years counting multiple {\it EUVE} observations, giving it long-term monitoring comparable to the early {\it ROSAT} flares.   With the first detection early in the \cha\ mission at $\simgreat700$ counts, the data quality in early epochs is comparable to more recent tidal flare examples such as SDSS J1201+30 reported by \cite{Saxton12} and 2XMMi J1847-63 reported by \cite{Lin11}.  Attempts to describe early \cha\ epochs produce a soft excess when fit with simple blackbody models, hinting at a more complicated scenario, possibly one described by different physical regions of the disrupted material.  There are also indications of additional structure beyond the much better power-law fit.  Additional modeling may be productive and should be compared to the flares of \cite{Saxton12} and \cite{Lin11}.  

One of the most significant new findings is the likelihood that the flare was produced by a MBH in a very small ($M_V\sim-14.7$) dwarf galaxy.  This conclusion would make it one of the least massive galaxies known to host a MBH, an order of magnitude smaller than either Henize 2-10 \citep{Reines11} or POX 52 \citep{Thornton08}, two of the smallest galaxies thought to host MBHs.  Given the small projected distance from the cluster core, we may suppose that WINGS J1348 could have had a previous interaction that stripped some fraction of the stellar material from a larger galaxy.

The {\it EUVE} identification poses a plausible solution to the unidentified {\it EUVE} transient found by \cite{BBK99}, and would make the WINGS J1348 flare the only TDF to be studied in the EUV $\simgreat0.02$~keV and $\simless0.1$~keV, below the \xmm\ sensitivity band in a regime critical to determining $L_{bol}$ in TDFs.  Given the lack of facilities in this wavelength regime, WINGS J1348 may also present the only confident EUV identification of a TDF for the foreseeable future.  Further analysis of {\it EUVE} may be possible with sufficient expertise, as the Short Wave spectrometer in particular has $\sim10$ per cent of the sensitivity of the Deep Survey camera, potentially allowing (for example) bright line spectroscopy or a more comprehensive determination of $L_{bol}$ during the 1998 March 27 {\it EUVE} epoch.  Any such lines from the TDF itself are likely to be too broad for detection given the limited number of source counts and high background \cite{SQ09}, but narrow emission lines may be temporarily excited in the surrounding interstellar medium.  The plausibility of alternate non-TDF theories of the flare's origin could also be examined in more detail.  Such analysis is well beyond the scope of this work.  The opportunity does, however, point to an additional potential scientific argument for future {\it EUV} missions \citep[in addition to those posed in][]{Kowalski10}.

The limits we present using 1800\;s of MagE spectroscopy and nondetections in most bands between the far ultraviolet and far infrared confidently exclude variability from persistent QSOs, NLS1s, and other bright, low-obscuration AGNs as explanations for the flare.  We also place strong constraints on rare extreme variability from Seyfert 2 galaxies such as from a disc state change or change in column density, although we cannot rule them out entirely.  Nondetections via repeated X-ray observations of A1795 provide evidence that the \cha\ and \xmm\ flare of WINGS J1348 does not arise from any obvious periodic accretion, as may be the case for HLX-1 \citep{Lasota11}.  

We also note that the detection of a decayed WINGS J1348 flare within $\sim56$~kpc of the cluster core in \xmm\ indicates impressive promise for cluster X-ray surveys for the purpose of identifying TDFs, even with a large ($\sim20$-arcsec) PSF.  The rate at which these flares occur $\sim10^5$~galaxy$^{-1}$~year$^{-1}$ may be insufficient to justify a dedicated monitoring program in itself, but a strong case may be made in support of a coordinated survey with multiple scientific goals.  Already several flares have been found associated with galaxy clusters \citep[][and this work]{Cappelluti09,Maksym10}, supporting the potential utility of such a program.  And the detection of a heavily decayed tidal flare even deep in the bright diffuse emission of the cluster core implies that the large \xmm\ PSF of \xmm\ or even {\it eROSITA} \citep{eRosita07} should be no barrier to such a program with careful choice of a low-energy filter.  The deep cluster observation program of {\it eROSITA} \citep{eRosita07}, due to launch by 2014, in particular holds significant potential in light of these results.  We also suggest X-ray observers of galaxy clusters using proposal-awarded time, that they endeavour to report flaring point sources in a timely manner, in order to better enable rapid follow-up of proprietary data.

We conclude that we have most likely detected a TDF from a dwarf galaxy in A1795.  Deeper optical spectroscopy will greatly assist in distinguishing between the competing explanations, and we will continue to seek such supporting observations.  Given the relatively small distance of A1795 and the time elapsed since disruption ($\sim15$ years), confident determination of these properties opens a wide range of potential future observational inquiries.  The properties of the host galaxy and \Mbh\ for the MBH are also of interest, particularly if WINGS J1348 is indeed a cluster member.  Deep high-resolution spectroscopy beyond the basic requirements of redshift determination and AGN limits would also allow determination of \Mbh\ via absorption line dispersion, or an upper limit if \Mbh\ is consistent with \cite{Lauer07}.  A spectral determination of \Mbh\ would determine not only whether the MBH is truly unusual for the size of its galaxy, but also help determine if the observed accretion is significantly super-Eddington.

If WINGS J1348 is a dwarf galaxy in A1795, it will be interesting to examine the galaxy morphology to better determine its evolutionary history and any evidence of recent interactions in the cluster environment.  Deep {\it Hubble} observations have the potential to detect extended low surface brightness features beyond the $\sim300$~pc optical core, and would allow spatial analysis not possible given contamination of the pre-existing WFPC2 images by cosmic ray artefacts.  Deep, high-resolution radio observations such as with VLBI or ALMA would allow investigation of host morphology and star formation history.  As the time of writing $\sim15$ years post-disruption, associated radio light echoes should be resolvable to $\sim3$~mas, and the expansion of any previously associated off-axis jet \citep[as per][]{VKF11,GM11,DeColle12} to $\sim0.3$~mas.  Despite being well past peak radio emission ($\sim1$ year), any associated jet may still be visible at $\sim$mJy sensitivity at $z\sim0.062$, potentially providing useful constraints on jet formation models and on the composition of the ISM within $\sim1$~pc of the MBH.  

\section*{Acknowledgments}

We thank the referee for numerous helpful comments which greatly improved the quality of this paper.

PM and MU acknowledge the support of a NASA ADP grant NNX08AJ35G.
We gratefully acknowledge the support of NASA for continued operations of
$HST$, \cha\ and their data archives, the ESA's support of \xmm\ and
its archives and the continued efforts of the Sloan Digital Sky
Survey.  This research has made use of the NASA/IPAC Extragalactic Database (NED) which is operated by the Jet Propulsion Laboratory, California Institute of Technology, under contract with the National Aeronautics and Space Administration.   This work is based [in part] on observations made with the Spitzer Space Telescope, which is operated by the Jet Propulsion Laboratory, California Institute of Technology under a contract with NASA.  We have made use of the ROSAT Data Archive of the Max-Planck-Institut f\"ur extraterrestrische Physik (MPE) at Garching, Germany. PM and MU thank Laura Klein for her assistance in data reduction.
PM thanks Jimmy Irwin and Bill Keel for helpful discussions, and Jeremy Drake for helpful instrumental discussions regarding {\it EUVE}.  

Funding for the SDSS and SDSS-II has been provided by the Alfred P. Sloan Foundation, the Participating Institutions, the National Science Foundation, the U.S. Department of Energy, the National Aeronautics and Space Administration, the Japanese Monbukagakusho, the Max Planck Society, and the Higher Education Funding Council for England. The SDSS Web Site is http://www.sdss.org/.

The SDSS is managed by the Astrophysical Research Consortium for the Participating Institutions. The Participating Institutions are the American Museum of Natural History, Astrophysical Institute Potsdam, University of Basel, University of Cambridge, Case Western Reserve University, University of Chicago, Drexel University, Fermilab, the Institute for Advanced Study, the Japan Participation Group, Johns Hopkins University, the Joint Institute for Nuclear Astrophysics, the Kavli Institute for Particle Astrophysics and Cosmology, the Korean Scientist Group, the Chinese Academy of Sciences (LAMOST), Los Alamos National Laboratory, the Max-Planck-Institute for Astronomy (MPIA), the Max-Planck-Institute for Astrophysics (MPA), New Mexico State University, Ohio State University, University of Pittsburgh, University of Portsmouth, Princeton University, the United States Naval Observatory, and the University of Washington. 

\bibliography{apj-jour,pete_tidal,biblio_mel_marc}
\bibliographystyle{mn2e}  

\bsp

\label{lastpage}

\end{document}